\documentclass[usenatbib,usedcolumn]{mnras}

\usepackage[T1]{fontenc}
\usepackage{ae,aecompl}

\usepackage{graphicx}	
\usepackage{amssymb}
\usepackage[fleqn]{amsmath}
\usepackage{array}
\usepackage{lmodern}

\newcommand{\epseri}{$\epsilon$ Eri}
\newcommand{\clean}{\texttt{clean}}
\newcommand{\ujypbm}{$\mu$Jy\,beam$^{-1}$}
\newcommand{\ujy}{$\mu$Jy}

\newcommand{\mSun}{\; {\rm M_\odot}}					%
\newcommand{\mJup}{\; {\rm M_{J}}}					%
\newcommand{\lSun}{\; {\rm L_\odot}}						%
\newcommand{\au}{\; {\rm au}}								%
\newcommand{\um}{\; {\rm \mu m}}						%
\newcommand{\myr}{\; {\rm Myr}}							%
\newcommand{\auPerMyr}{\; {\rm au \; Myr^{-1}}}		%
\newcommand{\kgPerMCubed}{\; {\rm kg \; m^{-3}}}	%
\newcommand{\percent}{\; {\rm per \; cent}}				%

\begin{document}

\title[The clumpy structure of $\epsilon$ Eri's debris disc]{The clumpy structure of \texorpdfstring{$\epsilon$}{epsilon} Eridani's debris disc revisited by ALMA}
\author[M. Booth et al.]{Mark Booth$^{1}$\thanks{E-mail: markbooth@cantab.net}, Tim D. Pearce$^{1}$, Alexander V. Krivov$^{1}$, Mark C. Wyatt$^{2}$, William R. F. Dent$^{3}$, \newauthor Antonio S. Hales$^{3,4}$, Jean-Fran\c{c}ois Lestrade$^{5}$, Fernando Cruz-S\'aenz de Miera$^{6}$, \newauthor Virginie C. Faramaz$^{7}$, Torsten L\"ohne$^{1}$ and Miguel Chavez-Dagostino$^{8}$\\
$^{1}$ Astrophysikalisches Institut und Universit\"atssternwarte, Friedrich-Schiller-Universit\"at Jena, Schillerg\"a\ss{}chen 2-3, D-07745 Jena, \\Germany  \\
$^{2}$ Institute of Astronomy, University of Cambridge, Madingley Road, Cambridge CB3 0HA, UK \\
$^{3}$ Joint ALMA Observatory, Alonso de C\'ordova 3107, Vitacura 763-0355, Santiago, Chile \\
$^{4}$ National Radio Astronomy Observatory, 520 Edgemont Road, Charlottesville, Virginia, 22903-2475, USA \\
$^{5}$ Observatoire de Paris, PSL Research University, CNRS, Sorbonne Universit\'es, UPMC, 61 Av. de l'Observatoire, F-75014 Paris, \\France \\
$^{6}$ Konkoly Observatory, Research Centre for Astronomy and Earth Sciences, E\"otv\"os Lor\'and Research Network (ELKH), \\ Konkoly-Thege Mikl\'os \'ut 15-17, 1121 Budapest, Hungary \\
$^{7}$ Steward Observatory, Department of Astronomy, University of Arizona, 933 N. Cherry Ave, Tucson, AZ 85721, USA \\
$^{8}$ Instituto Nacional de Astrof\'isica Optica y Electr\'onica Luis Enrique Erro \#1, CP 72840, Tonantzintla, Puebla, M\'exico
}

\makeatletter
\hypersetup{
	pdfauthor=\@shortauthor,
	pdftitle=\@title,
	bookmarksnumbered=true
	}
\makeatother

\date{Accepted 2023 March 23. Received 2023 March 7; in original form 2022 April 12}
\pubyear{2023}
\volume{{\rm in press}}

\maketitle

\begin{abstract}
$\epsilon$ Eridani is the closest star to our Sun known to host a debris disc. Prior observations in the (sub-)millimetre regime have potentially detected clumpy structure in the disc and attributed this to interactions with an (as yet) undetected planet. However, the prior observations were unable to distinguish between structure in the disc and background confusion. Here we present the first ALMA image of the entire disc, which has a resolution of 1.6\arcsec$\times$1.2{\arcsec}. We clearly detect the star, the main belt and two point sources. The resolution and sensitivity of this data allow us to clearly distinguish background galaxies (that show up as point sources) from the disc emission. We show that the two point sources are consistent with background galaxies. After taking account of these, we find that resolved residuals are still present in the main belt, including two clumps with a $>3\sigma$ significance -- one to the east of the star and the other to the northwest. We perform $n$-body simulations to demonstrate that a migrating planet can form structures similar to those observed by trapping planetesimals in resonances. We find that the observed features can be reproduced by a migrating planet trapping planetesimals in the 2:1 mean motion resonance and the symmetry of the most prominent clumps means that the planet should have a position angle of either ${\sim10^\circ}$ or ${\sim190^\circ}$. Observations over multiple epochs are necessary to test whether the observed features rotate around the star.
\end{abstract}


\section{Introduction}
\epseri{} is the closest star to our Sun known to host a debris disc\footnote{There is a claim that Proxima, the closest star to the Sun, hosts a multi-component debris disc \citep{anglada17}, but \citet{macgregor18} demonstrated that the unresolved excess is a result of stellar flares and Chittidi et al. (in prep.) showed that the structure claimed as a resolved belt is not seen in more recent observations.}. This was one of the first debris discs detected \citep{gillett86} and one of the first resolved \citep{chini91, greaves98}. The disc has been detected from mid-infrared to millimetre wavelengths \citep{gillett86,chini91, greaves98,schutz04, backman09,greaves14a,lestrade15,macgregor15,chavez16,booth17,holland17,su17,ertel20}. The bulk of the far-infrared and millimetre emission comes from a narrow (11~au wide)
belt (hereafter referred to as the main belt) that is close to face-on situated around 70~au from the star \citep{booth17}. The belt's fractional width of $\Delta R/R=0.17$ makes it one of the narrowest discs known \citep{lestrade15}, with a fractional width comparable to that of the classical Kuiper belt \citep[e.g.][]{gladman01}.
The warm emission close to the star is dominated by dust from either one broad disc extending from around 3 to 20~au or two belts at around 2 and 8~au \citep{su17}.

Starting with the original Submillimetre Common-User Bolometer Array (SCUBA) image of the disc taken at the James Clerk Maxwell Telescope (JCMT) \citep{greaves98}, there have been claims of clumps and asymmetries in the main belt. \citet{greaves98} clearly identified a bright peak in the belt to the east of the star. They also noted peaks in the northeast, northwest and southwest parts of the belt, although found that these only appear in half of their maps. These features were recovered in follow-up observations from SCUBA by \citet{greaves05}. They compared the locations of the features and proposed that the northeast peak, northwest peak and an arc of material in the south were real features of the disc as they showed a common proper motion and hints of an anti-clockwise rotation, whilst bright points to the east and southwest did not show this common proper motion and so are likely background galaxies. However, they also noted that the $\sim3\arcsec$ positional uncertainty meant that these conclusions were not definitive. Observations using the Max-Planck-Millimeter-Bolometer (MAMBO) data from the Institut de Radioastronomie Millim\'etrique (IRAM) 30m telescope \citep{lestrade15} showed a similar azimuthal profile to the SCUBA data, with peaks in the northeast, northwest and southwest parts of the belt consistent with the SCUBA data, although these peaks were all detected at a signal to noise ratio (SNR) of less than 3. Observations using the \emph{Herschel Space Observatory} \citep{greaves14a} did not show any signs of azimuthal variations except for a flux asymmetry between the north and south sides of the belt at 160$\um$, 250$\um$ and 350$\um$ (that the authors attributed to pericentre glow). Observations using AzTEC on the Large Millimeter Telescope (LMT) \citep{chavez16} also showed this flux asymmetry. Some less pronounced azimuthal variations were also seen, although the data were found to be consistent with a smooth belt \citep{chavez16}. Both the \emph{Herschel} and AzTEC observations detected a number of galaxies to the east of the star \citep{greaves14a, chavez16}, one of which is in a position consistent with the bright source detected in the first SCUBA observations i.e. it has not followed the westward proper motion of the star. Observations with SCUBA-2 \citep[also at the JCMT;][]{holland17} at 850$\um$ show significant substructure with bright clumps in the southeast and northwest of the disc. To summarise, the brightest source in the original SCUBA data has since been identified as a background galaxy that is now to the east of the disc; a bright spot or arc is significantly detected in the south-southeast part of the disc in multiple datasets, but has not yet been identified as either a disc feature or background; and other low significance structures (SNR$\approx2-3$ higher than the rest of the belt) around the belt have been identified in similar locations in multiple datasets, but their existence has yet to be conclusively proven.
\begin{table*}%
\begin{tabular}{c>{\centering}p{1.7cm}cccc}
Observation Start & Time on Source (min) & Antennas & PWV (mm) & Bandpass Calibrator & Phase Calibrator \\
\hline
2019-12-16 00:31:05 & 49.59 & 42 & 1.11 & J0423-0120 & J0315-1031  \\
2019-12-17 03:52:57 & 49.59 & 43 & 1.04 & J0423-0120 & J0348-1610  \\
2019-12-18 23:51:06 & 49.59 & 42 & 1.26 & J0334-4008 & J0315-1031  \\
2019-12-22 02:10:33 & 49.59 & 44 & 0.92 & J0423-0120 & J0315-1031  \\
2019-12-23 03:23:39 & 49.59 & 41 & 2.01 & J0423-0120 & J0348-1610 \\
\end{tabular}
\caption{Details of the observations. PWV stands for precipitable water vapour.}
\label{tobs}
\end{table*}

The reported detection of clumps in \epseri{} %
inspired theoretical work to understand the origin of these structures. By considering our own Solar System, \citet{liou99} investigated how a fraction of the small grains produced through collisions in the Kuiper belt and then dragged towards the Sun due to Poynting-Robertson (P-R) and stellar wind drag would end up being captured into the resonances of Neptune and thus produce resonant clumps -- a process that could also be occurring around other stars. This process was then considered as an explanation for the clumpy emission seen in extrasolar Kuiper belts \citep{ozernoy00,quillen02,deller05}. However, the effect of drag forces is expected to be weak in systems with bright debris discs \citep{wyatt05}. If the dust remains in the same location as its parent planetesimals, then an alternative explanation can also be made via comparison with the Kuiper belt, where many Kuiper belt objects are trapped in mean-motion resonance (MMR) with Neptune. The presence of these objects on resonant orbits is due to the migration of Neptune \citep[e.g.][]{malhotra95,levison03}. By analogy, the presence of resonant clumps in extrasolar systems can be explained by dust produced through collisions of planetesimals that have been trapped by a migrating planet in the system \citep{wyatt03, wyatt06, krivov07}.

One planet has so far been detected in this system \citep{hatzes00}. By combining radial velocity, astrometry and direct imaging, \citet{llop21} determined that this planet (known as \epseri{} b) has a mass of 0.7$\pm$0.1~$\mJup$, semi-major axis of 3.52$\pm$0.04~au and eccentricity of 0.07$^{+0.08}_{-0.05}$.
The small semi-major axis and low eccentricity of this planet mean that it is unlikely to be responsible for clumps in the main belt. So far, no planets have been detected close to the main belt. Direct imaging observations place a limit of ${\sim1\mJup}$ on planets between $\sim$25~au and the main belt \citep{janson15}. Until future observatories improve on these limits, the properties of outer planets in the disc can only be inferred from their influence on the disc. The Atacama Large Millimeter/submillimeter Array (ALMA) provides an excellent opportunity to do this as its high resolution and sensitivity enable us to clearly distinguish background galaxies from disc features and accurately map the disc features.

In this paper we present new ALMA observations of the \epseri{} system. The observations are described in Section \ref{sobs}, in which we detect three point sources along with the main belt. The point sources and main belt are analysed in Sections \ref{spoint} and \ref{sdisc}. We demonstrate how clumpy structure in the belt can be explained by a migrating planet in Section \ref{sclumps}. We discuss emission close to the star in Section \ref{sinner}. 
We present our conclusions in Section \ref{sconc}.

\section{Observations}
\label{sobs}

\epseri{} was observed by ALMA in cycle 7 for the project 2019.1.00696.S (PI: M. Booth). The data were taken in band 6 (1.29~mm) between 16th and 23rd December 2019 (see Table \ref{tobs}). The total time on source was 4.7 hours. At this wavelength, the full width at half maximum (FWHM) of the primary beam is 22\arcsec{} and so one pointing is not enough to cover the entire belt. In order to cover the entire belt most efficiently with an approximately constant sensitivity, we set up the observations with six pointings in a hexagonal pattern around the disc, each separated by 18$\arcsec$.

\begin{figure*}
	\centering
	\includegraphics[width=0.30\textwidth]{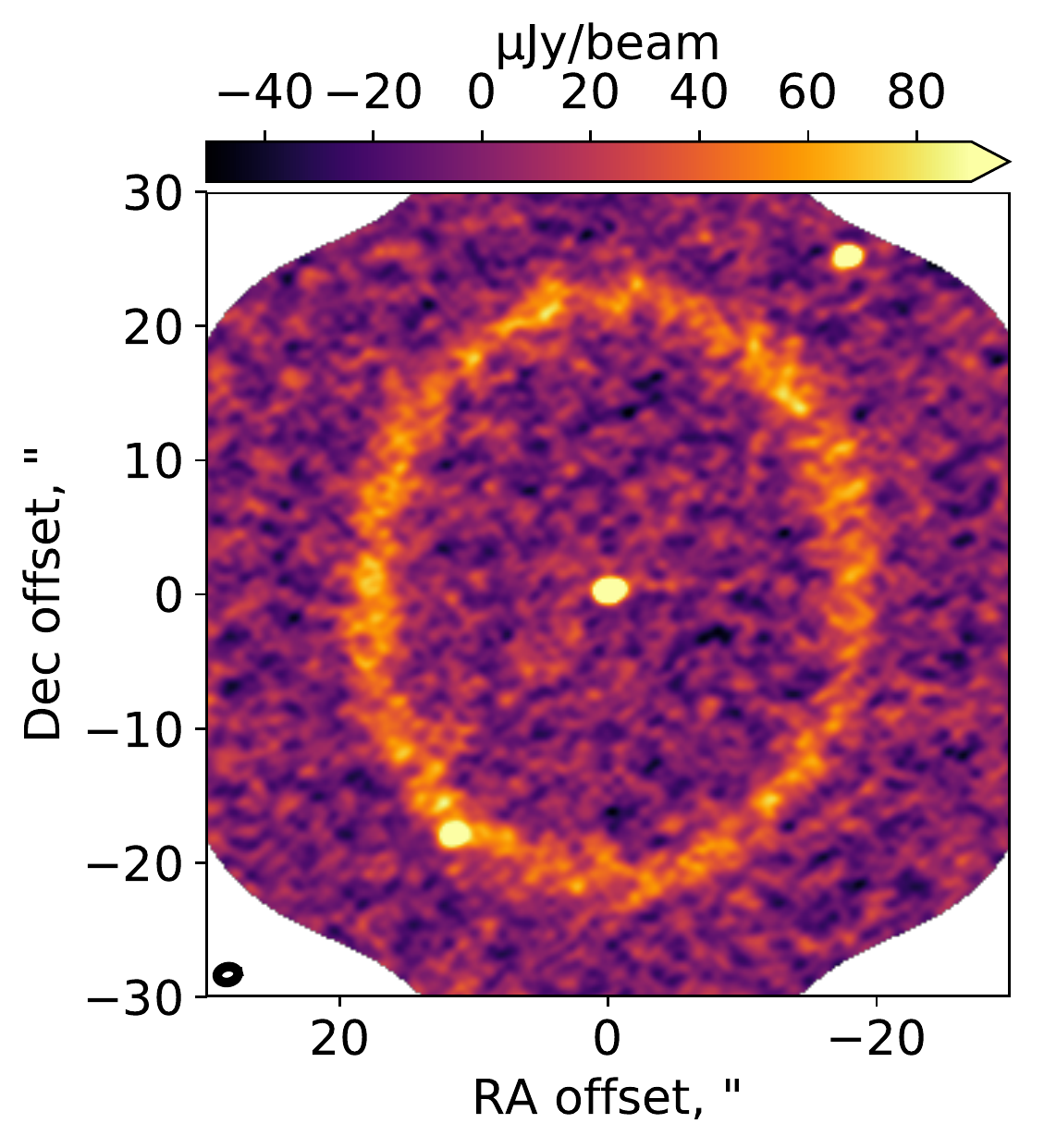}
	\includegraphics[width=0.30\textwidth]{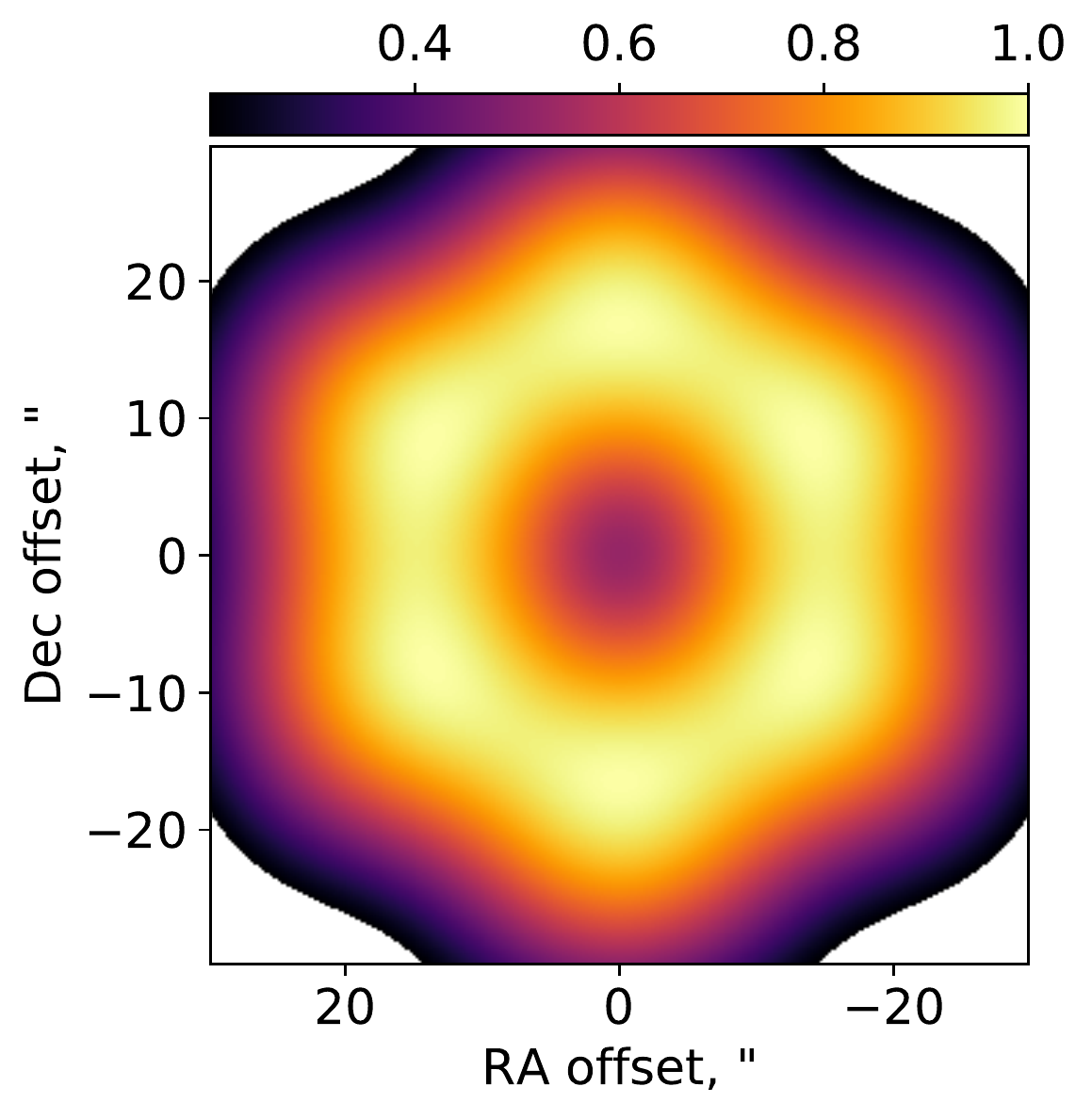}
	\includegraphics[width=0.30\textwidth]{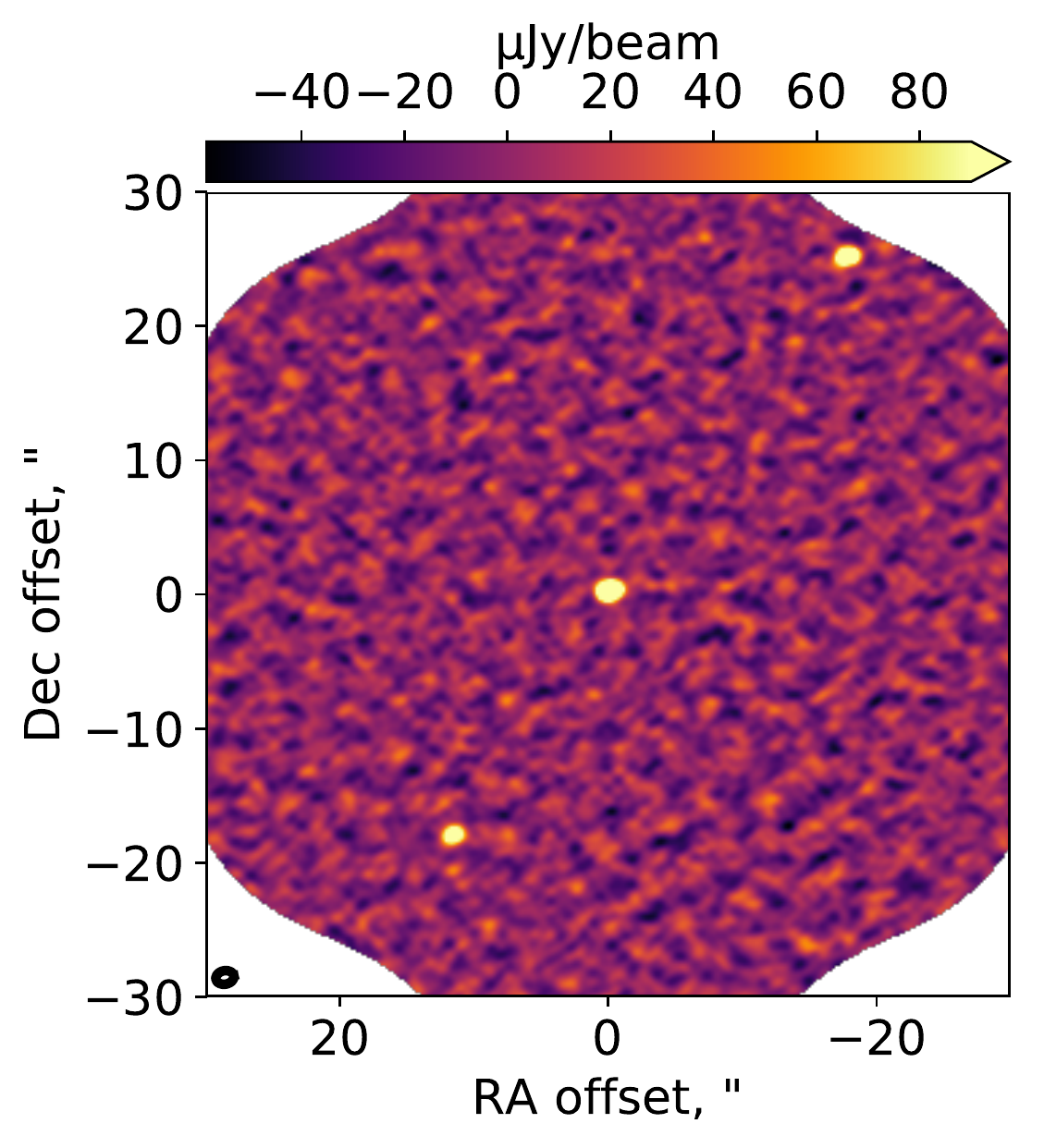}
	\caption{\emph{Left:} \clean{} image of \epseri{} using natural weighting. This is shown without correcting for the primary beam (as are all images in this paper), which means that the flux density shown is lower than the real flux density by a factor given in the middle image (i.e. dividing this image by the middle image produces a primary beam corrected image). We show the images without correcting for the primary beam so that the noise level is constant across the image and we can more easily discern visually real emission from noise. \emph{Middle:} Primary beam level. We observed the system with six pointings spread around the main belt. At the location of the star, the primary beam level is 53\%. \emph{Right:} \clean{} image created by including only baselines longer than 40k$\lambda$. This shows the three point sources present in the observations.}
	\label{fclean}
\end{figure*}

The observations were taken whilst the array was in a compact configuration with baselines between 15 and 331~m. The correlator was configured such that one spectral window, centred on the CO(2-1) line at 230.538~GHz, had 3840 channels of width 0.5 MHz (0.6~kms$^{-1}$) and the other three spectral windows, centred at 232, 245 and 247~GHz, had 128 channels of width 16~MHz (21~km\,s$^{-1}$). For the continuum imaging we used the full bandwidth of all four channels adding up to 7.875~GHz. The data were reduced in \textsc{casa} \citep{mcmullin07} version 5.6.1-8 using the standard observatory calibration, which includes water vapour radiometer correction, system temperature, complex gain calibration and flagging.

The left panel of \autoref{fclean} shows the \clean{} image, created using natural weighting and multi-frequency synthesis. The original \clean{} algorithm \citep{hogbom74} determines the sky intensity distribution by iteratively fitting the Fourier transform of point sources to the visibilities. In order to better fit extended sources, \citet{cornwell08} developed a multiscale version of the \clean{} algorithm that supplements the point sources with circular sources of a range of scales. Due to the large angular scale of the disc compared to the beam, we used this multiscale \clean{} \citep{cornwell08} to create our image.
For this, we set the scales to 0\arcsec{}, 1.4\arcsec{} and 7\arcsec{}. For the \clean{} mask, we used an elliptical mask covering the disc and a circular mask covering the source northwest of the disc. We set the cleaning threshold to 25~\ujypbm, which was reached after 1273 iterations. The resulting synthesised beam has a size 1.6\arcsec$\times$1.2{\arcsec} with a position angle of 102\degr{} E of N. The sensitivity of the final image is $\sigma=14$~\ujypbm, determined by measuring the RMS in a number of regions both interior and exterior to the main belt. The image has not been corrected for the primary beam so that the sensitivity remains constant across the image (the primary beam is also shown in the middle panel of \autoref{fclean}). The image shows three point sources as well as the main belt of the system.

\begin{table*}%
\begin{tabular}{cccc}
Identifier & Right ascension & Declination & Peak flux density, \ujypbm \\ 
\hline
Central & 3h\;32m\;54.521s$\pm$0.006 & -9d\;27m\;29.49s$\pm$0.07 & 1020$\pm$50 \\
NW & 3h\;32m\;53.325s$\pm$0.007 & -9d\;27m\;04.61s$\pm$0.08 & 990$\pm$60 \\
SE & 3h\;32m\;55.307s$\pm$0.013 & -9d\;27m\;47.66s$\pm$0.15 & 270$\pm$30 \\
\end{tabular}
\caption{Positions and flux densities of the three point sources based on Gaussian fits to the primary beam corrected long baseline image. The positional uncertainties are dominated by the astrometric accuracy \protect\citep[see section 10.5.2][]{alma22}. An extra 5\% calibration uncertainty has been added in quadrature to the uncertainty of the flux densities \protect\citep[see section 10.2.6][]{alma22}.}
\label{tps}
\end{table*}

To check for CO $J$=2-1 emission, we also used \clean{} to create a data cube of the channels around the expected radial velocity of the star \citep[16.376$\pm$0.0019~km\,s$^{-1}$][]{soubiran18}. No CO line emission was detected even when integrating around the belt (between radii of 63 and 76~au) and adjusting for the expected orbital velocity \citep[i.e., using the spectro-spatial filtering method of][]{matra15}. From this we find the $3\sigma$ upper limit on an unresolved emission line to be $5.0\times10^{-22}$~W\,m$^{-2}$ \citep[using Equation A1 of][]{booth19}. Based on a model of gas production through the collisional cascade, \citet{kral17} predicted that this system would have a CO $J$=2-1 level of $1.1\times10^{-24}$~W\,m$^{-2}$. Our upper limit is, therefore, consistent with this model and much deeper observations will be necessary to detect any gas in this system.

\begin{figure*}
	\centering
	\includegraphics[width=0.98\textwidth]{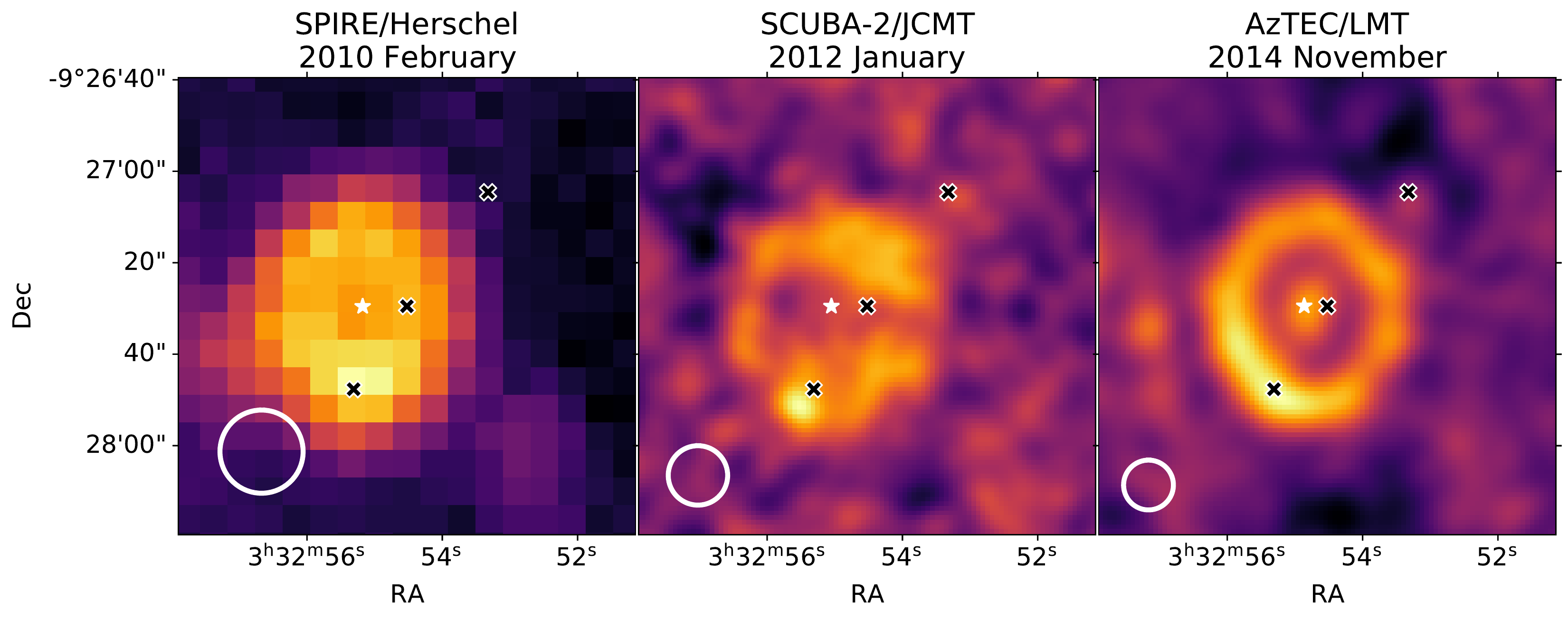}
	\caption{Point sources from cycle 7 ALMA data (shown by crosses) overplotted on the images from \emph{Herschel}/SPIRE 250~$\mu$m \citep{greaves14a}, SCUBA-2/JCMT \citep{holland17} and AzTEC/LMT \citep{chavez16} without any correction for the proper motion. The beam size for each dataset is indicated by the white circle. The position of the star at the time of each observation is indicated by the white star. This is clearly distinct from the central cross due to the high proper motion of the star. The positions of the northwest and southeast point sources, however, are consistent with the locations of bright sources in the previous datasets and have not moved with the star demonstrating that these must be background sources.}
	\label{fbg}
\end{figure*}

\section{Analysis of point sources}
\label{spoint}
Three point sources are visible in the image. In order to isolate these from the disc emission, we create a \clean{} image with only the long baseline data (specifically we selected all baselines greater than 40~k$\lambda$ as the disc is only detectable at shorter baselines). This image is shown in the right panel of \autoref{fclean}. In order to determine the flux density, a correction for the effect of the primary beam must be applied to account for the fall-off in sensitivity away from the phase centre of the pointings. Here we used the standard \textsc{casa} assumption of an Airy disc \citep[see section 3.2][]{alma22}. 
After correcting the image for the primary beam, we then use \textsc{casa}'s Gaussian fitting function to determine the location and flux density of the sources (see \autoref{tps}). For the flux density we add a systematic uncertainty of 5\% in quadrature to the measured uncertainty based on the expected flux calibration uncertainty at this wavelength \citep[see section 10.2.6][]{alma22}. In addition, a systematic uncertainty due to the uncertainties in the primary beam of around 1\% is expected for primary beam levels below 50\% (Dirk Petry, private communication), such as where the star and northwest point source are located. However, as this is negligible compared to the flux calibration uncertainty \citep[see also][]{kundert17}, this is not taken into account. For the positional accuracy we use Equation 10.7 from \citet{alma22}
\begin{equation}
 \Delta\theta=\mathrm{FWHM}\times(0.9\times\mathrm{SNR})^{-1}
 \label{ealmaposunc}
\end{equation}
noting that this is valid up to an SNR of 20 and that the postional accuracies may be a factor of two or more worse than this depending on the conditions at the time of the observations.

The image is centred on the expected position of the star at the time of the observations ($\alpha$\;=\;3h\;32m\;54.519s, \mbox{$\delta$\;=\;-9d\;27m\;29.50s}). The central peak is consistent with this location within the uncertainties and so we conclude that this is a detection of the star. Note here that the image we are working with is created combining data taken over a week. 
The star has a high proper motion with a right ascension component of $\mu_{\alpha\star}=-974.76\pm0.16\;{\rm mas\;yr^{-1}}$ and declination component of $\mu_\delta=20.88\pm0.12\;{\rm mas\;yr^{-1}}$ \citep{gaia18} and so it will have moved westward by 0.019\arcsec{} between the first and last observation. Since this difference is roughly equivalent to the positional uncertainty, we can safely ignore it in the anaylsis. 

The flux density measured for the star is slightly higher than the 820$\pm$70\ujy{} measured in the cycle 2 dataset \citep{booth17}, but considering their uncertainties they are consistent at the 2$\sigma$ level. In addition, a slight difference is to be expected due to a slight difference in wavelength -- the cycle 2 dataset has a central wavelength of 1.34~mm, whereas the cycle 7 dataset presented here has a central wavelength of 1.29~mm. At the shorter wavelength, the photospheric emission will be higher, but the star is also known to have significant chromospheric and coronal emission \citep{lestrade15,macgregor15,booth17,bastian18,suresh20,mohan21} and it is not yet fully understood how this varies with wavelength. 

In \autoref{fbg} we compare the locations of the point sources (marked by black crosses) to prior observations from the Spectral and Photometric Imaging REceiver (SPIRE) instrument on \emph{Herschel} \citep[wavelength of 250~$\mu$m, angular resolution of 18.2\arcsec][]{greaves14a}, SCUBA-2/JCMT \citep[wavelength of 850~$\mu$m, angular resolution of 15\arcsec;][]{holland17} and AzTEC/LMT \citep[wavelength of 1.1~mm, angular resolution of 10.9\arcsec][]{chavez16}. Each observation also shows the location of the star at the time of the observation (marked by a white star), although only in the AzTEC data is the star actually detected.

The point source to the northwest is far beyond the belt and so is unlikely to be related to the \epseri{} system. It appears at the same right ascension and declination as peaks in the AzTEC and SCUBA-2 data, although these are both detected at a SNR$<3\sigma$. The point source to the southeast is coincident with the belt and so it is harder to distinguish from the belt in the prior observations due to the low resolution of those observations. The SPIRE and AzTEC data both show a brightness asymmetry that is brightest at this point. There is a bright ($\sim7\sigma$) point in the SCUBA-2 data that is offset from the southeast source by $\sim2.5\arcsec{}$ in both right ascension and declination. \citet{ivison07} showed that for SCUBA-2 the RMS uncertainties of the right ascension and declination of a point source can be expressed as 
\begin{equation}
 \Delta\alpha=\Delta\delta=0.6\times\mathrm{FWHM}\times{\mathrm{SNR}}^{-1}
\end{equation}
(where SNR means the signal to noise ratio). In addition to this, we need to add in quadrature the pointing uncertainty of the telescope, which is $\sim2\arcsec{}$ for JCMT\footnote{\url{https://www.eaobservatory.org/jcmt/about-jcmt/}}. This means that we expect the positional uncertainty of the SCUBA-2 bright point to be $\sim2.4\arcsec{}$ and so it is consistent with the location of the southeast source.
From this we conclude that the southeast source does not share a common proper motion with the star and so must also be a background source. 
We also conclude that both the northwest and southeast sources are most likely to be galaxies given their point like nature -- submillimetre galaxies are typically observed to have sizes $<1\arcsec$ \citep[e.g.][]{ikarashi15,fujimoto16,gomez22} and so will be unresolved in our observations.
This conclusion is also consistent with the fact that both sources are only detected at long wavelengths. The fact that the northwest source is only significantly detected at 1.3~mm, whereas the southeast source is detected at wavelengths between 250$\um$ and 1.3~mm suggests that the northeast source has a higher redshift.

\section{Analysis of the disc}
\label{sdisc}
As an interferometer, ALMA measures visibilities at a number of baselines corresponding to the separation between pairs of antennas. These visibilities represent the Fourier transform of the on sky intensity image. However, since an interferometer has a finite number and limited range of baselines, there are gaps in the coverage of the Fourier plane and the Fourier transform of the visibilities produces what is known as a `dirty' image, which includes various interferometric artifacts. This is particularly problematic for faint, extended emission, as is the case for \epseri{} debris disc. In order to refine the reconstructed image, one must determine an accurate model of the emission to fill in the gaps in the observation's coverage of the Fourier plane.

One commonly used method for this is \clean{}, which was used for our `quick-look' image shown in Figure \ref{fclean}. In Section \ref{sclean} we will describe the limitations of this method for our data. Another commonly used method is forward modelling using a parametric model, whereby a simple model of the emission with a number of free parameters is created for a range of parameters and the most likely model is determined. In Section \ref{sfm} we show how this is useful for determining the form of the disc, although it has limitations for structures in the disc.

\subsection{\clean{} model}
\label{sclean}
\begin{figure*}
	\centering
	\includegraphics[width=0.48\textwidth]{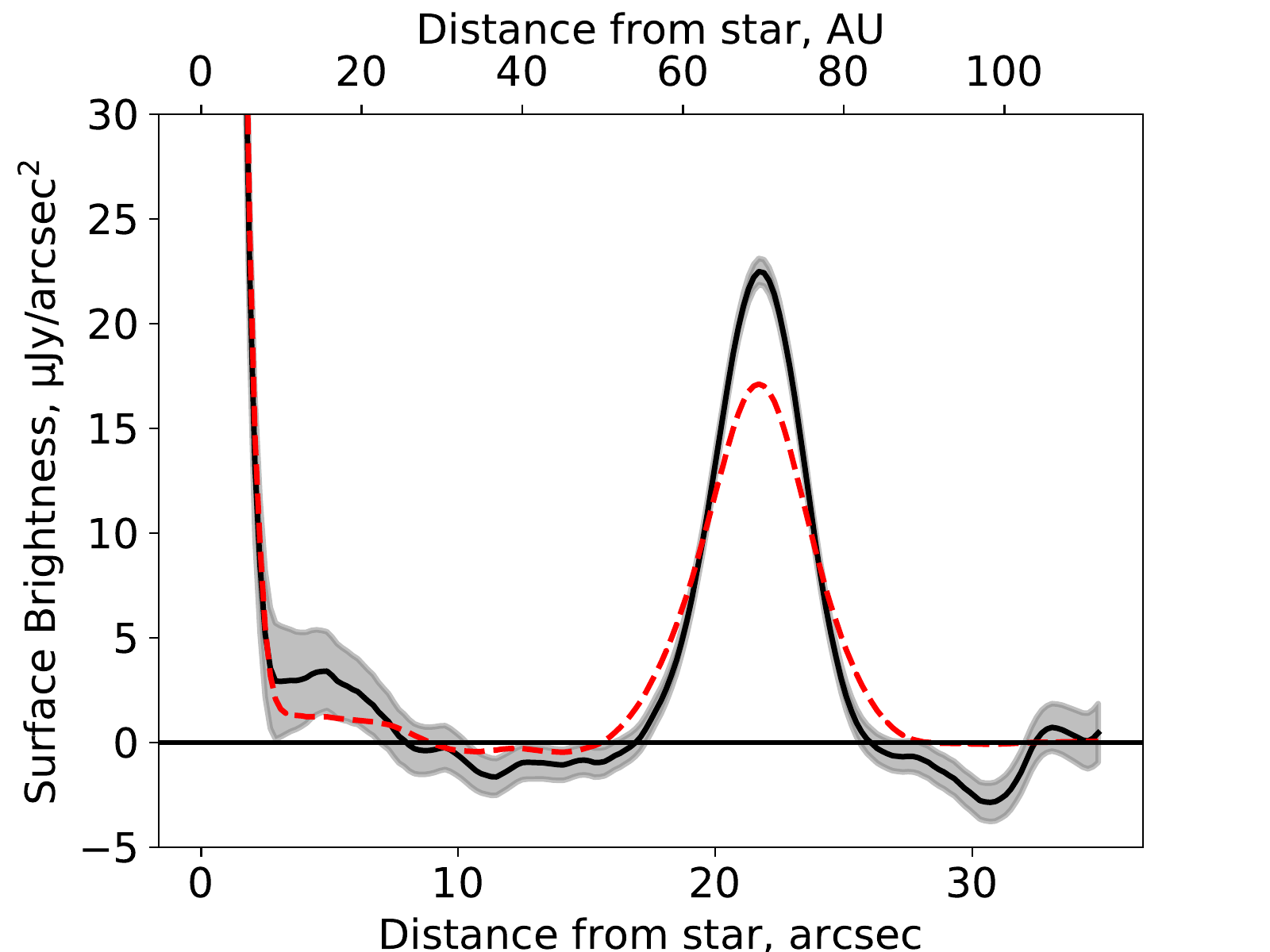}
	\includegraphics[width=0.48\textwidth]{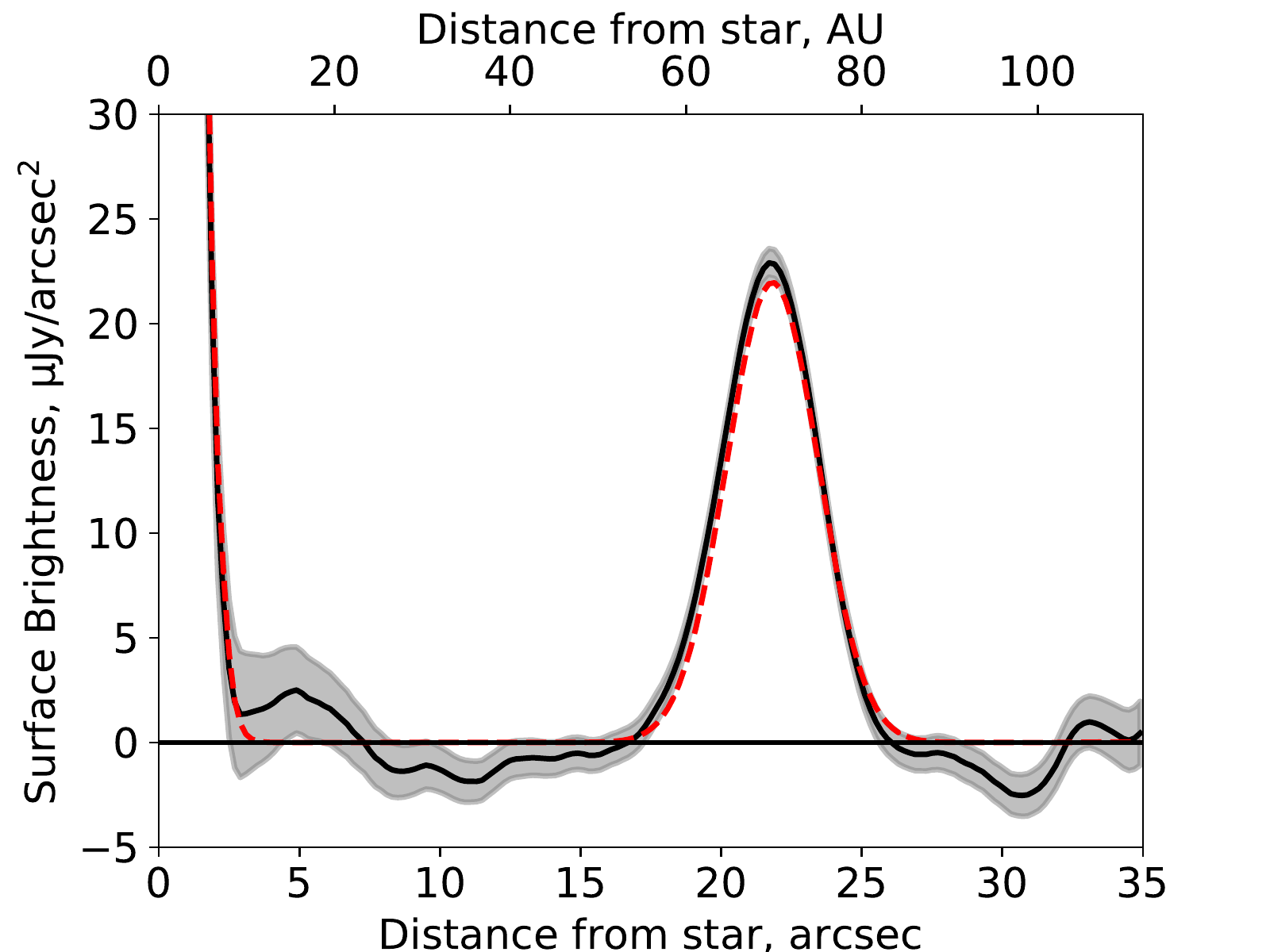}
	\caption{Deprojected surface brightness distributions of the result of the \clean{} algorithm (left) and forward modelling (right) after correcting for the primary beam. Both cases show the model convolved with the \clean{} beam both before (red dashed line) and after (solid black line, grey region shows the standard error of the weighted mean) residuals are added.}
	\label{fcleanprof}
\end{figure*}
In Section \ref{sobs} we used multiscale \clean{} \citep{cornwell08} for our `quick look' \clean{} image. Note that what is referred to as the \clean{} image is the \clean{} model convolved with the \clean{} beam -- produced by fitting a two dimensional Gaussian to the Fourier transform of the sampling function -- and added to the residuals. In order to test how good the \clean{} model is, we can take the images from before and after the residuals are added, deproject them \citep[assuming the inclination and position angle from][]{booth17} and azimuthally average. This comparison is shown in \autoref{fcleanprof}. From this we can see that the peak in the main belt emission is at $\sim$70~au, as expected from \citet{booth17}. However, we notice that the \clean{} algorithm does not manage to account for all of the flux in the main belt, i.e., there is significant residual flux which has not been `cleaned out', seen in the left panel of \autoref{fcleanprof}. This difference cannot be explained by a lack of cleaning sufficiently deep as the threshold we set for the \clean{} iterations is roughly twice that of the final RMS noise (see Section \ref{sobs}). Choosing a lower threshold than this will only result in an increased likelihood of fitting to noise rather than emission. Another possibility is that the difference between the \clean{} image and the \clean{} model is due to the difference in units between the convolved model and the residuals. Both are typically quoted in units of Jy\,beam$^{-1}$, however the beams are different. The model has been convolved with the \clean{} beam whereas the residuals are in terms of the dirty beam and so adding the two together can result in the residuals dominating. This effect was first noted by \citet{jorsater95} and has since become known as the JvM effect. \citet{czekala21} showed how the effect can be calculated from the ratio of the volume of the \clean{} beam to the volume of the dirty beam. In our case we find that this ratio is 0.98 and so has a negligible impact on the \clean{} image. Therefore we conclude that in order to accurately model the emission we need to start with a more realistic model of the source that we can fit simultaneously to all the emission.

\subsection{Forward modelling}
\label{sfm}
Next we test whether modelling the main belt with a parametric ring model and finding a best fit through forward modelling can better account for the emission than the multiscale \clean{} model. 
For our parametric model we assume a point source to represent the star as well as any unresolved flux (e.g. from dust within a few au of the star) and a circular ring with a Gaussian radial profile to represent the main belt. The system is also known to have one or two belts between 3 and 20~au \citep{backman09,greaves14a,su17}. These have so far only been detected at mid-infrared wavelengths and we do not see evidence for them in our data (Figure \ref{fclean}), therefore we do not include them in our model. The two background sources are also modelled by point sources fixed to the results from \autoref{tps}. We therefore have six free parameters: the flux density of the central point source ($F_\mathrm{cen}$), the flux density of the belt ($F_\mathrm{belt}$), the distance of the peak of the belt emission from the star ($R_\mathrm{mid}$), the FWHM of the belt ($\Delta R$), the inclination of the belt with respect to the sky plane ($I$) and the position angle measured anti-clockwise from north ($\Omega$).

Comparison with the observations can either be done in the image plane or in the Fourier plane. When modelling the cycle 2 data for this system (taken at the same wavelength with the same resolution but with only one pointing and fewer antennas), \citet{booth17} compared both methods and found that there was no significant difference in their results. For simplicity, therefore, we fit to the data in the image plane \citep[as described in][]{booth17}. This involves creating a dirty image (i.e. an image that has not been processed by the \clean{} algorithm) using natural weights, an image of the point spread function (PSF, also referred to as the dirty beam) and an image of the primary beam. A Markov Chain Monte Carlo (MCMC) routine is set up using \textsc{emcee} \citep{foreman13} with 120 walkers and run for 1000 timesteps. For each set of parameters, a model image is created, multiplied by the primary beam and convolved with the PSF. This is then compared to the dirty image in order to calculate a likelihood given by
\begin{equation}
\ln \mathcal{L}=-\chi^2/2
\end{equation}
where
\begin{equation}
 \chi^2=\sum^{N}_{i=1} \left(\frac{O_i-M_i}{S_{ncr}\sigma}\right)^2,
\end{equation}
and $O_i$ and $M_i$ represent pixels in the observed image and model image respectively. $N$ is the total number of pixels used in the calculation. We note that the primary beam image we use is a Gaussian model of the beam. The actual primary beam may deviate slightly from this, which adds extra uncertainty to pixels far from the phase centre. We, therefore, only include pixels where the primary beam power is $>$20\% of the peak in the primary beam. $S_{ncr}$ is the noise correlation ratio equivalent to the square root of the number of pixels per beam. This is required because the high resolution of the image compared to the beam size has the side effect of introducing correlated noise. Uniform priors were used for all parameters.

\begin{table}
 \begin{tabular}{cc}
 Parameter & Best fit \\
 \hline
 $R_\mathrm{mid}$, au & $69.6\pm0.3$ \\ %
 $\Delta R$, au & $10.5\pm0.5$ \\ %
 $F_\mathrm{belt}$, mJy & $10.3\pm0.4$ \\ %
 $I$, \degr{} & $33.7\pm0.5$ \\ %
 $\Omega$, \degr{} & $-1.1\pm1.0$ \\ %
 $F_\mathrm{cen}$, mJy & $0.95\pm0.03$ \\ %
\end{tabular}
\caption{MCMC results for the Gaussian ring model. The uncertainties given are the 16th and 84th percentiles and do not include calibration uncertainties.}
\label{tres}
\end{table}

\begin{figure*}
	\centering
	\includegraphics[width=0.98\textwidth]{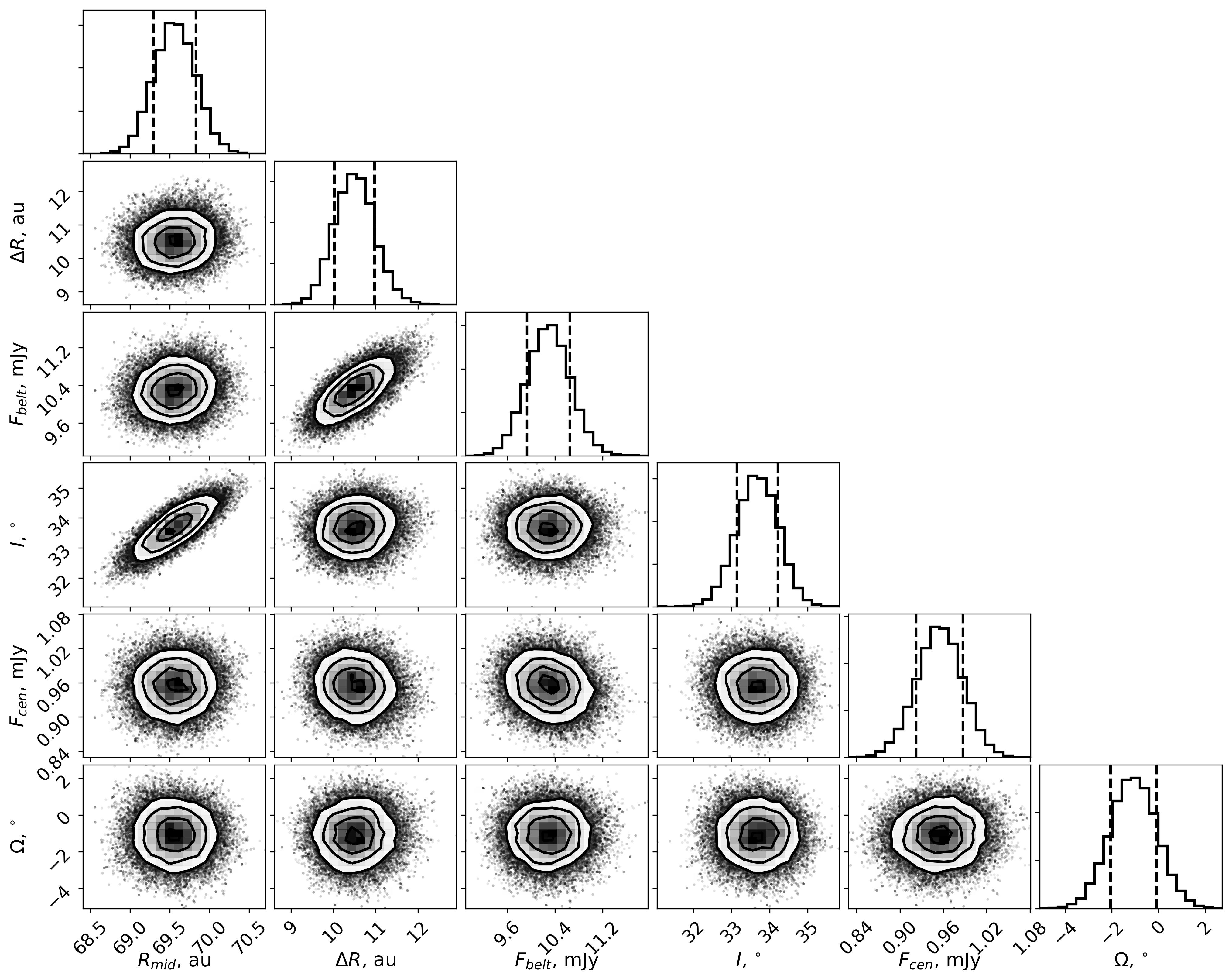}
	\caption{Marginalised posterior probability distributions for the free parameters in our model. Vertical dashed lines show the 16th and 84th percentiles. }
	\label{fcorner}
\end{figure*}

\begin{figure*}
	\centering
	\includegraphics[width=0.32\textwidth]{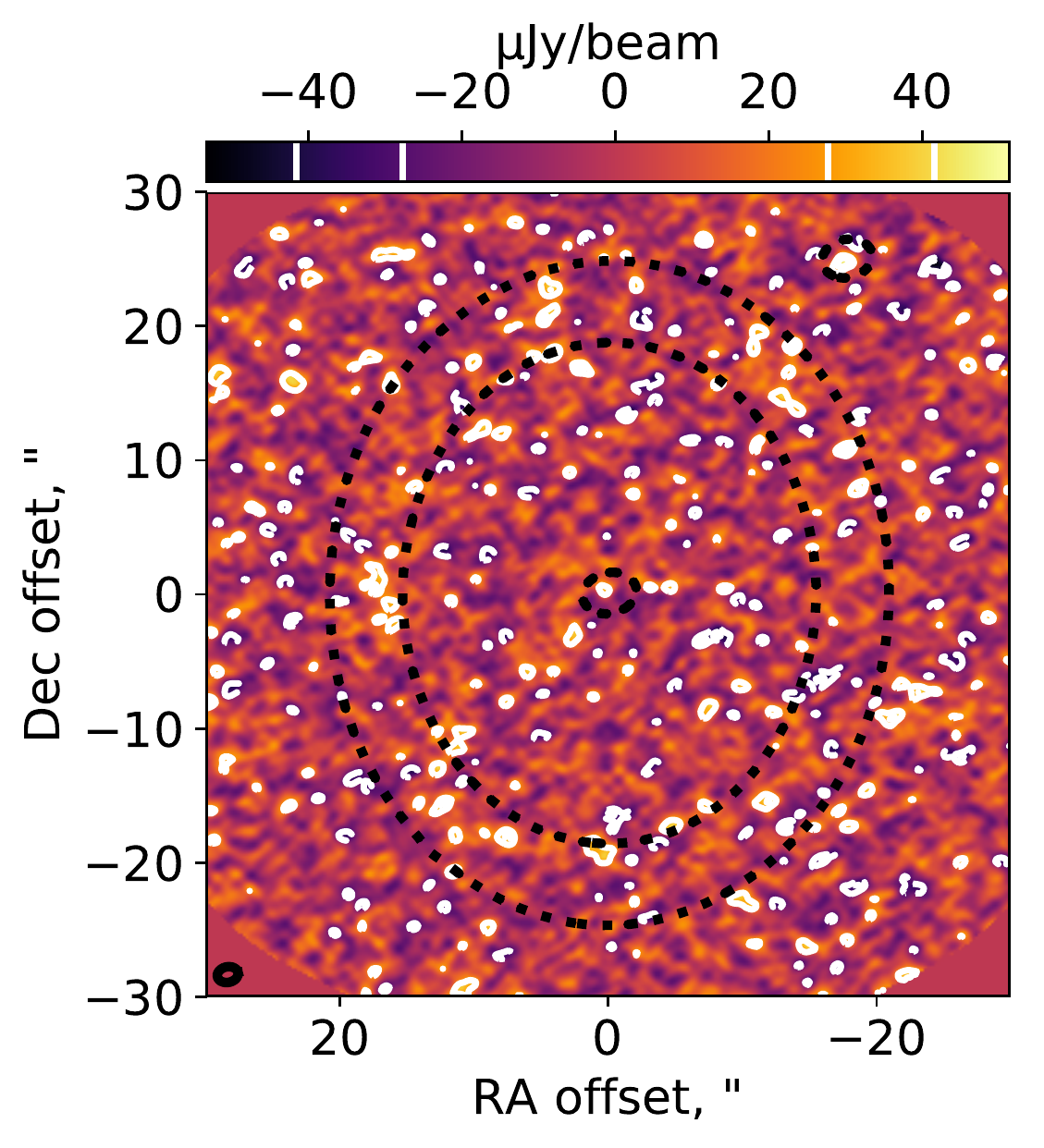}
	\includegraphics[width=0.32\textwidth]{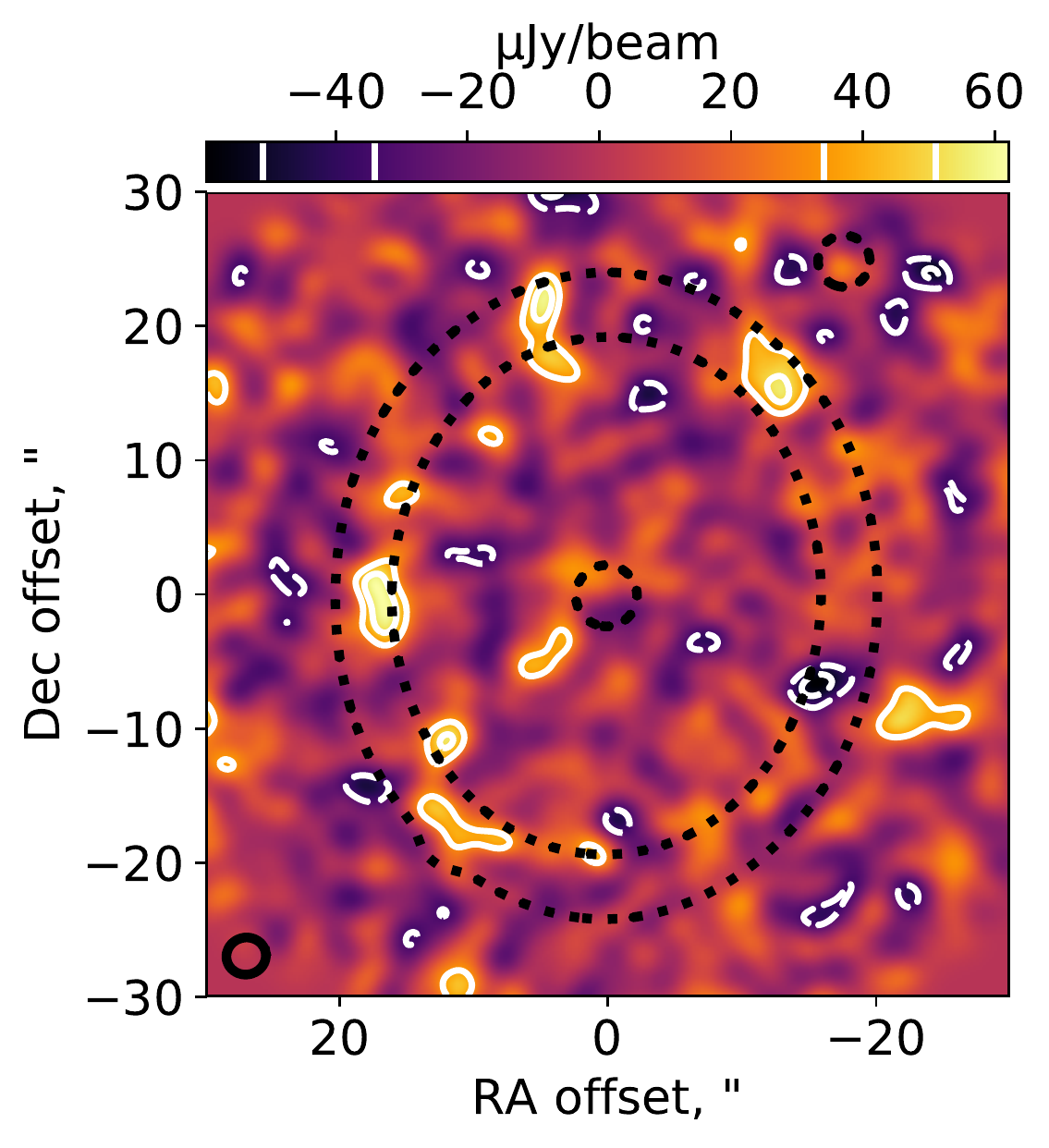}
	\includegraphics[width=0.32\textwidth]{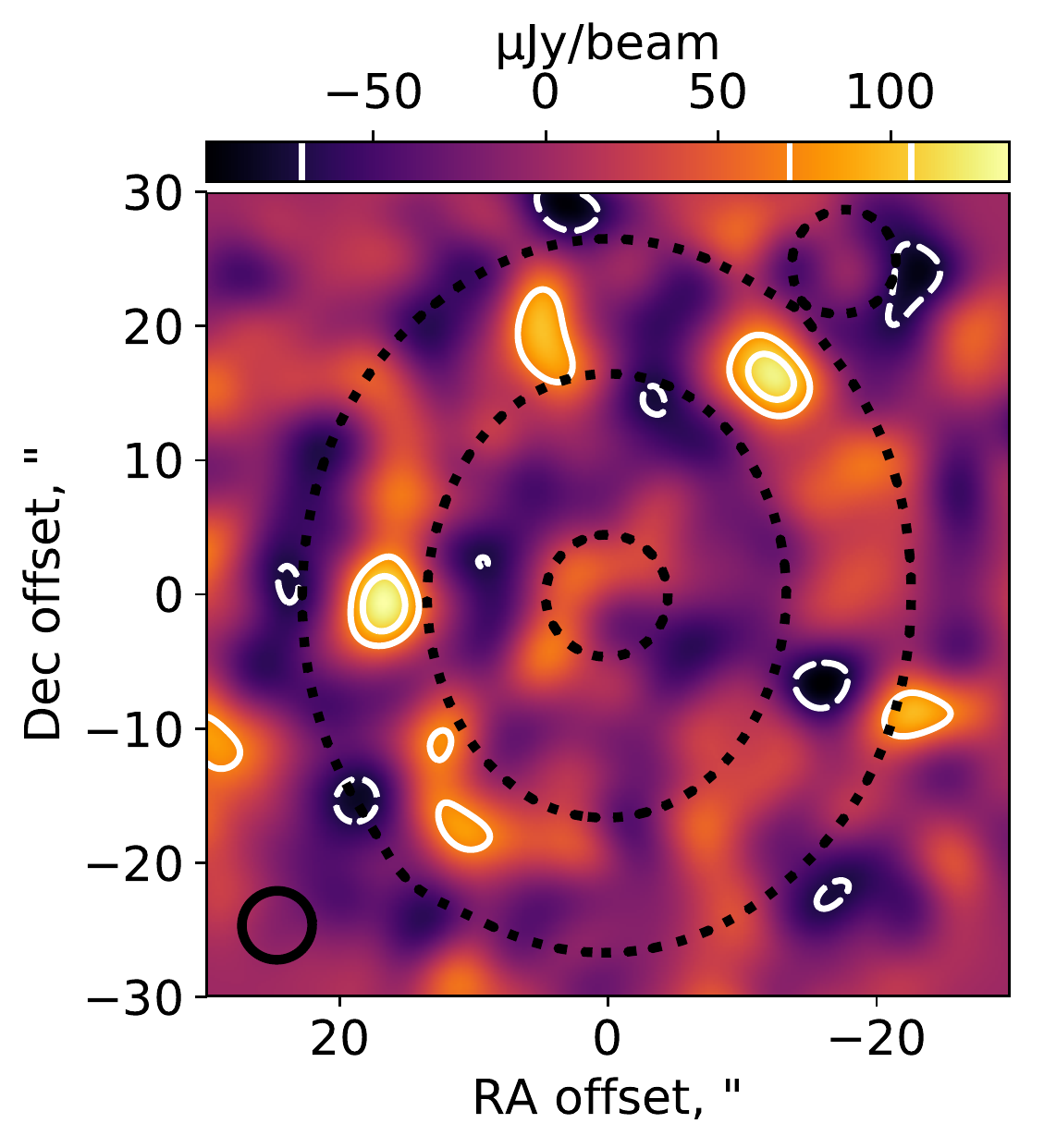}
	\caption{Residuals after subtracting the best fit MCMC model using natural weighting (left), after smoothing with a 2.5\arcsec{} Gaussian (middle) and after smoothing with a 5\arcsec{} Gaussian (right). White contours show the $\pm$2,3$\sigma$ residuals. Black dashed line shows the MCMC model.}
	\label{fgres}
\end{figure*}

The MCMC converges after about 250 timesteps. We then use the samples from the remaining 750 timesteps to infer the disc parameters and their uncertainties (determined by the 16th and 84th percentiles), which we show in \autoref{tres}. These results are consistent (at the 1$\sigma$ level) with the same model from \citet{booth17}, with the exception of the central flux density, which is slightly higher here (as already discussed in Section \ref{spoint}), although still consistent at the 2$\sigma$ level. The marginalised posterior probability distributions for each parameter and correlations between parameters are shown in Figure \ref{fcorner}. We find that all parameters have Gaussian posterior probability distributions. We also see that there is a correlation between $R_\mathrm{mid}$ and $I$ and another correlation between $\Delta R$ and $F_\mathrm{belt}$. 

\begin{figure}
	\centering
	\includegraphics[width=0.48\textwidth]{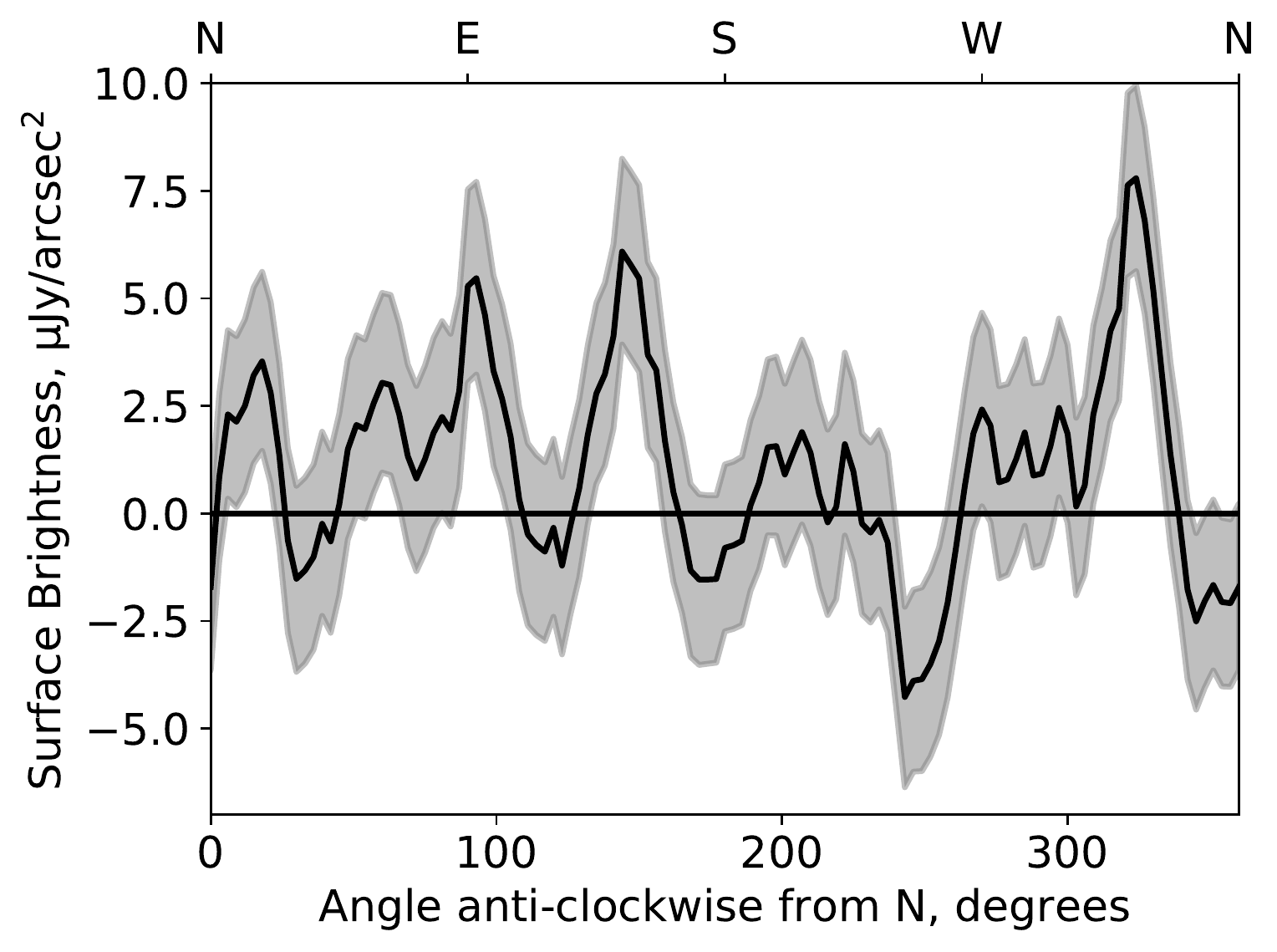}
	\caption{Azimuthal profile of the residuals left by our best-fit model (see the right plot of \autoref{fgres}) after correcting for the primary beam and deprojecting the disc based on the inclination and position angle found in Section \ref{sdisc}. The surface brightness is calculated from the region between 64 and 77 au and within an azimuthal range of 20$\degr$.}
	\label{faz1}
\end{figure}

The residuals of the best fit model are shown in the left plot of Figure \ref{fgres} and the surface brightness radial distribution is shown in the right plot of Figure \ref{fcleanprof}. This simple Gaussian model fits the radial profile of the main belt well, never deviating from the observed radial distribution by more than 3$\sigma$. Given the azimuthal variations described by previous authors, we also show an azimuthal profile of the residuals in Figure \ref{faz1}. This is created by correcting for the primary beam and taking the weighted mean of the surface brightness between stellar distances of 64 and 77 au and within an azimuthal range of 20$\degr$, calculated for every 5$\degr$ around the belt. As for the radial profile, the uncertainties are given by the standard error of the weighted mean. Here we see some variations in the surface brightness with a 3.6$\sigma$ peak in the northwest part of the disc.

\begin{figure*}
	\centering
	\includegraphics[width=0.30\textwidth]{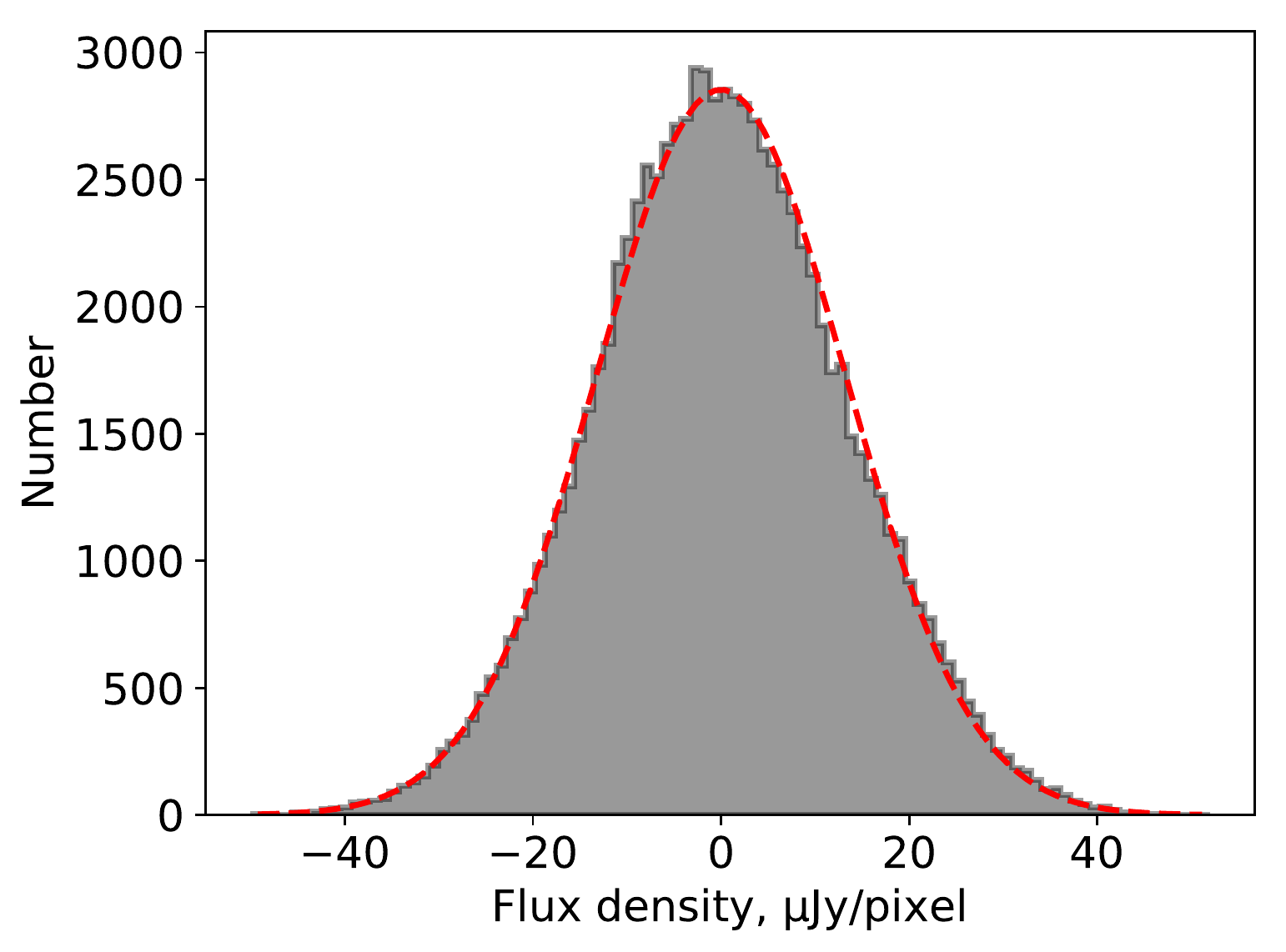}
	\includegraphics[width=0.30\textwidth]{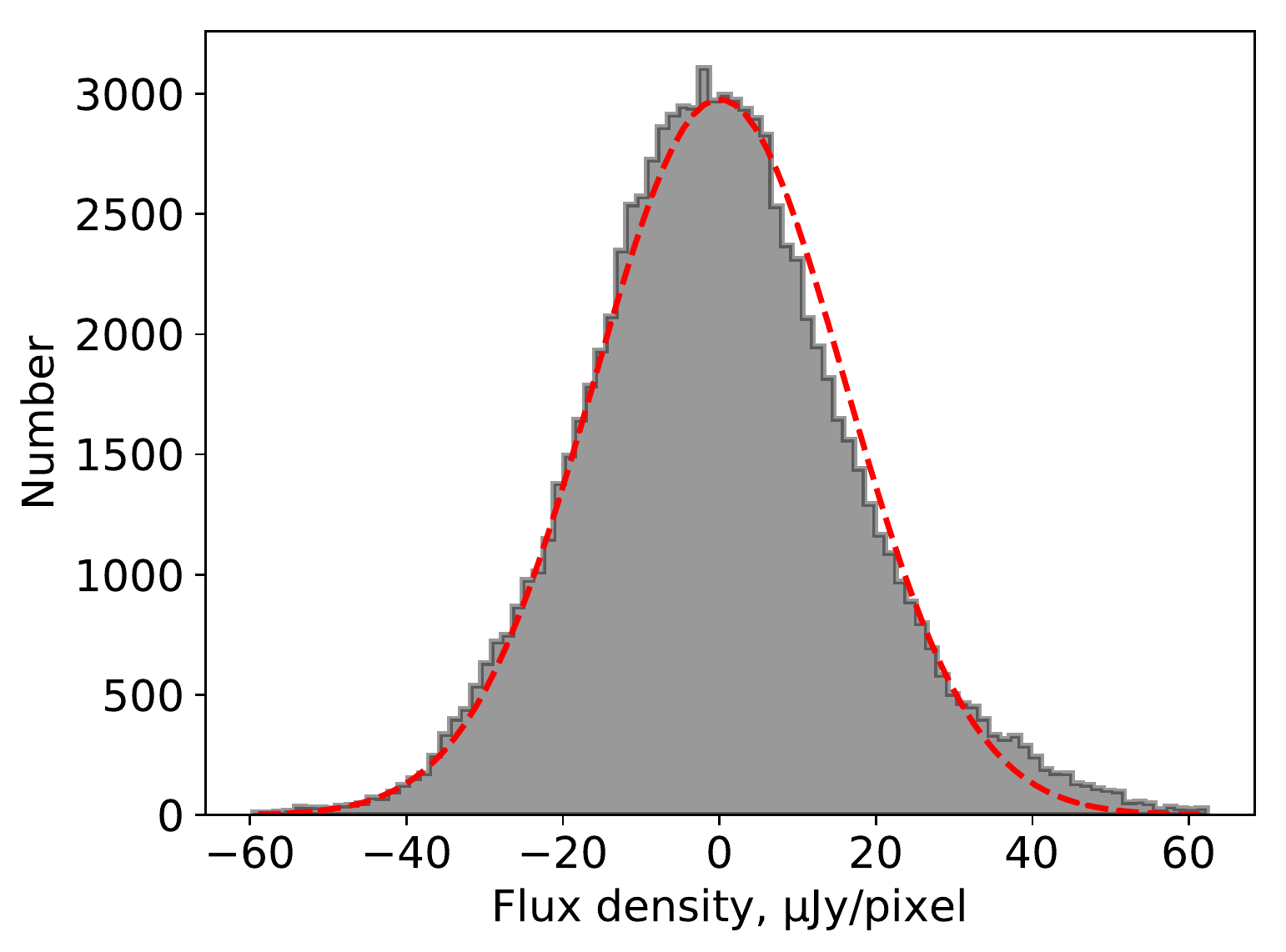}
	\includegraphics[width=0.30\textwidth]{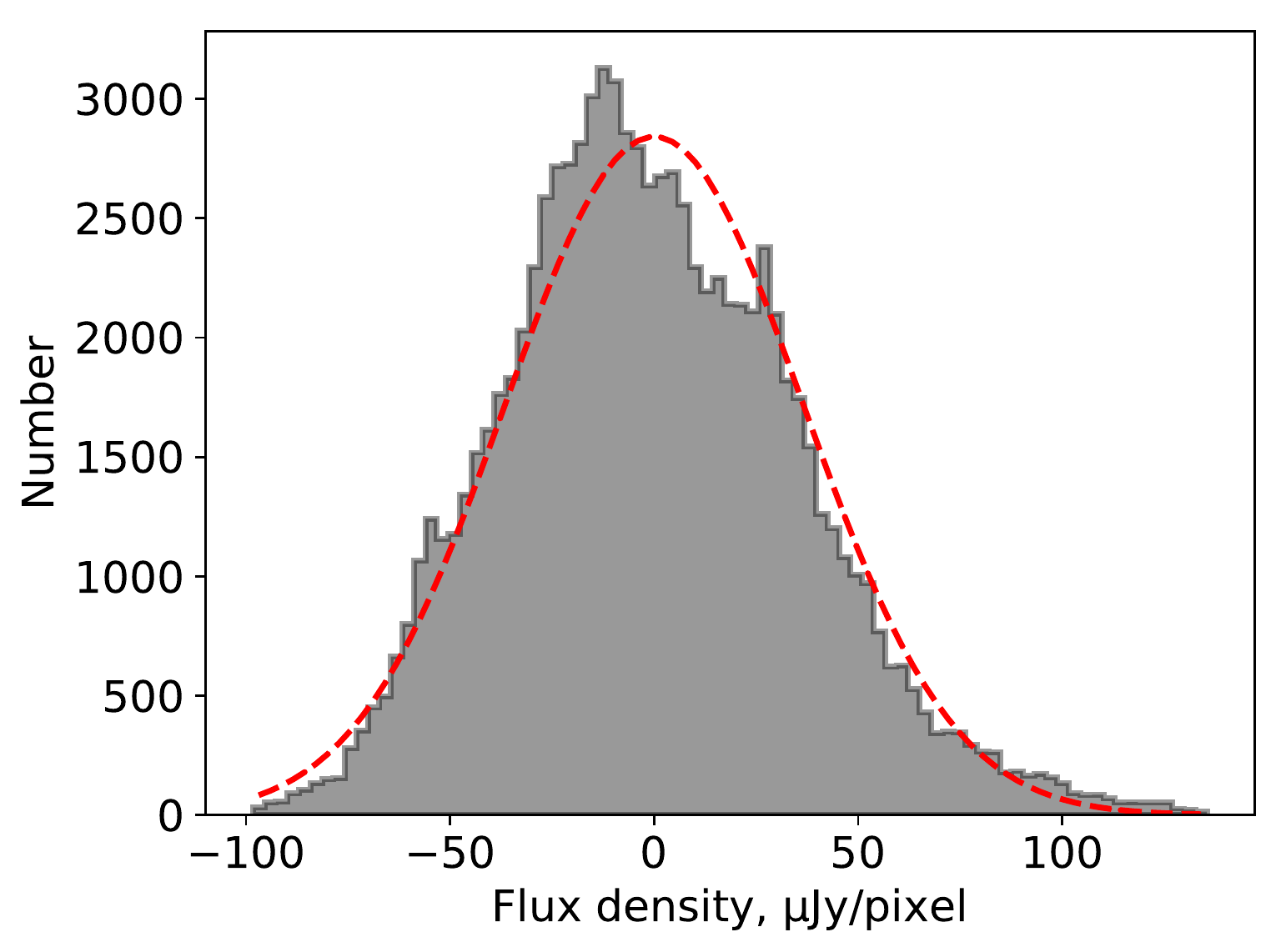} \\
	\includegraphics[width=0.30\textwidth]{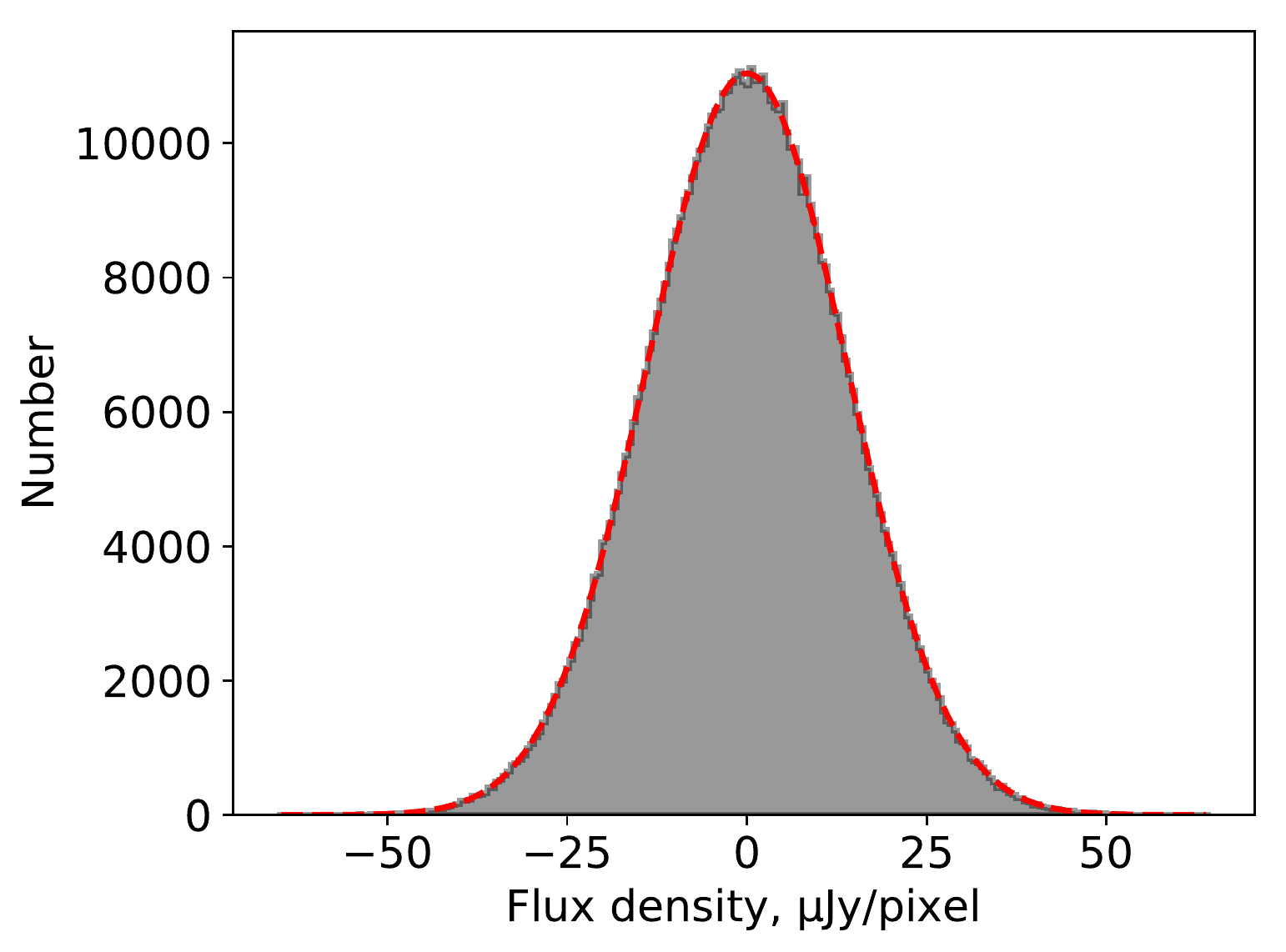}
	\includegraphics[width=0.30\textwidth]{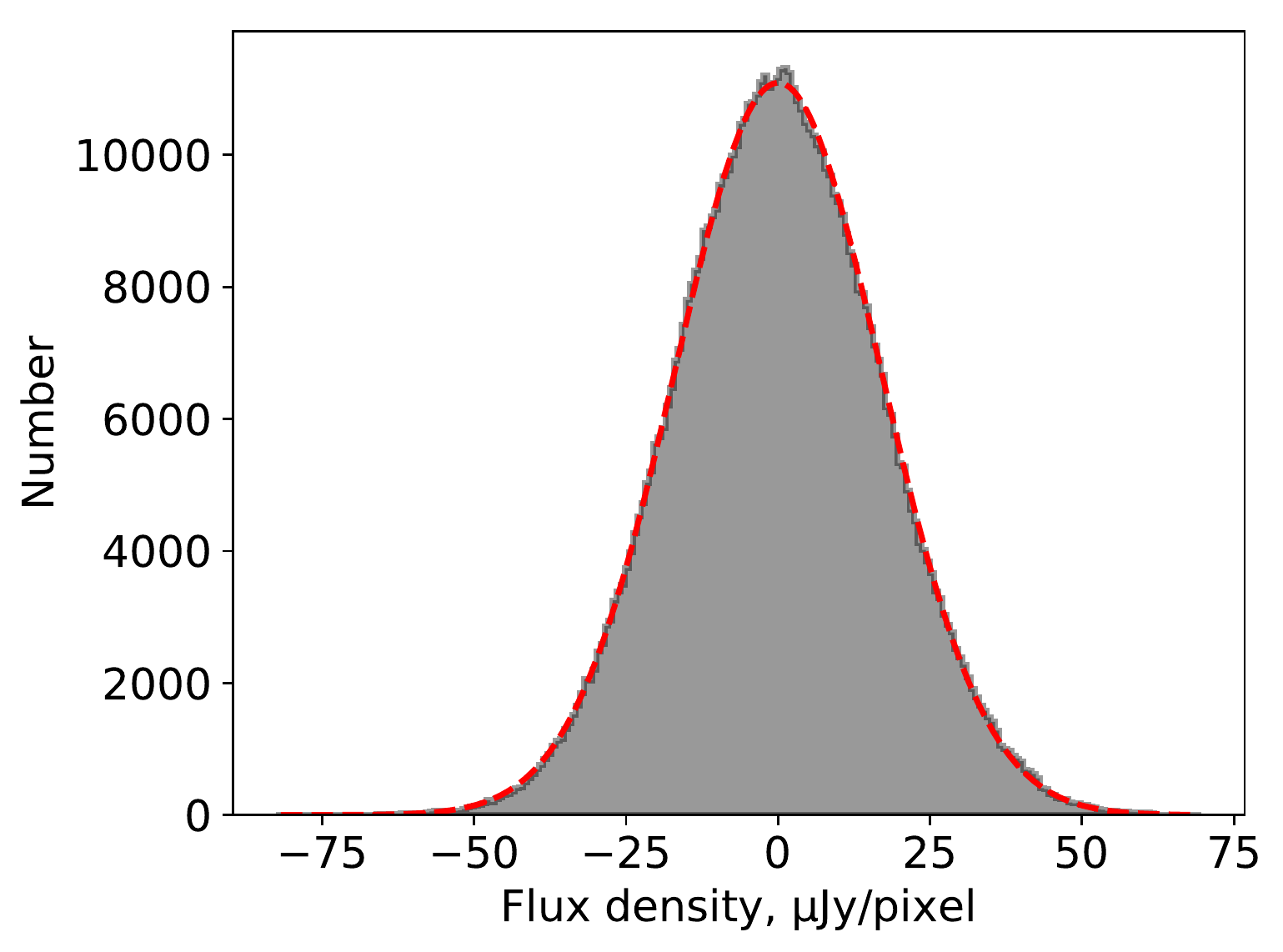}
	\includegraphics[width=0.30\textwidth]{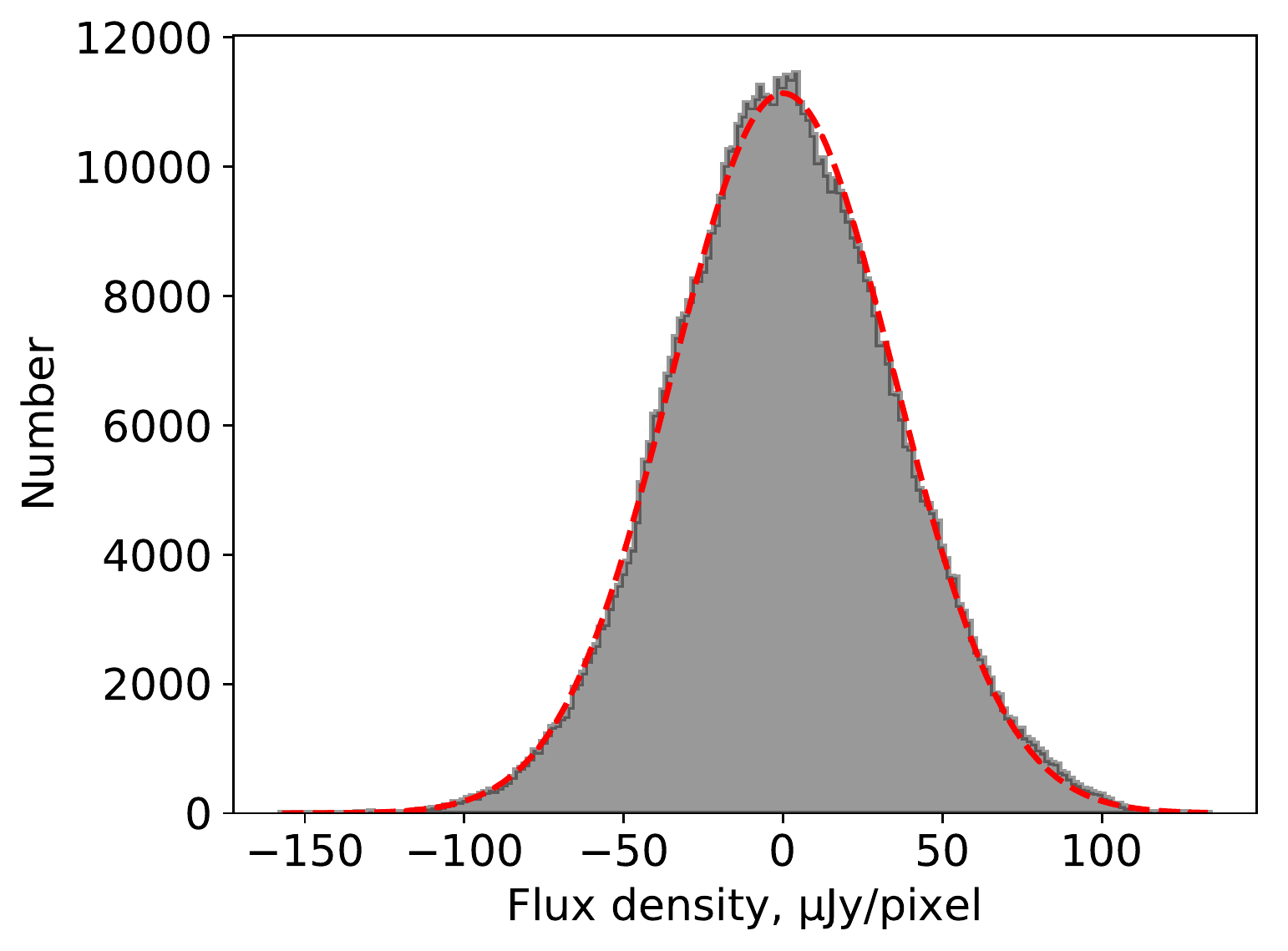}
	\caption{\emph{Top:} Histograms of the residual image (\emph{left}) and the residuals smoothed with a 2.5\arcsec{} Gaussian (\emph{middle}) and a 5\arcsec{} Gaussian (\emph{right}). As we can see, smoothing the residuals emphasises the non-Gaussianity of the residuals, demonstrating that the smooth model does not fully account for the detected emission. \emph{Bottom:} Histograms of the map created at a direction offset from the disc in order to measure the noise without influence from the signal (\emph{left}) and the versions of this map once it has been smoothed with a 2.5\arcsec{} Gaussian (\emph{middle}) and a 5\arcsec{} Gaussian (\emph{right}). In all plots a Gaussian with a mean of zero is fitted to the histograms and shown by the red dashed line.}
	\label{fhists}
\end{figure*}

The left plot of \autoref{fgres} shows mostly $2\sigma$ residuals, with little structure. However, this is a high resolution image and so we may be resolving out features of the disc, especially given the significant residual in the northwest seen in the azimuthal profile. In order to find out if we detect any clumps in this data, we need to degrade the resolution to improve the SNR. This can be achieved by smoothing the image with a Gaussian function. The middle and right plots of Figure \ref{fgres} show the result of smoothing with a Gaussian that has a FWHM of 2.5\arcsec{} and 5\arcsec{} respectively. Note that the actual beam size of the resultant image is not exactly the same as the applied Gaussian -- here we find them to be 2.7\arcsec$\times$2.8{\arcsec} and 5.1\arcsec$\times$5.2{\arcsec} respectively. As the disc emission and its subsequent interferometric artefacts fill the image, accurately determining the background noise from the image becomes more difficult for lower resolutions as there are fewer independent, disc-free beams. This issue is highlighted in the top row of Figure \ref{fhists} in which a histogram of the pixels is compared to a Gaussian with a mean of zero. The offsets from the Gaussian profile show these interferometric artefacts that are due to the presence of residual emission in the data. In order to determine that actual noise level of the data so that we can accurately quantify the significance of the residuals, we use \textsc{casa} to create an image offset from the source that shows only background noise, as in \citet{booth17}. This image cannot be created using the mosaic gridder and so a single pointing is selected. Due to the spacing between our pointings (selected to provide a constant sensitivity around the belt), the theoretical noise level in an image created using a single pointing is equivalent to the theoretical noise level of the mosaic image created using all pointings. In our case, we create the noise only image with phasecentre that is offset by 3\arcmin{} from the star -- an offset that is found to be large enough to avoid any influence from the detected emission but not so large that the beam size is drastically changed. We first test this process with the original unsmoothed data and confirm that we measure $\sigma=14$~\ujypbm{}, as in Section \ref{sobs}. Using this method for the smoothed images, we measure $\sigma=17$~\ujypbm{} in the 2.5\arcsec{} smoothed map and $\sigma=35$~\ujypbm{} in the 5\arcsec{} smoothed map.

\begin{figure}
	\centering
	\includegraphics[width=0.48\textwidth]{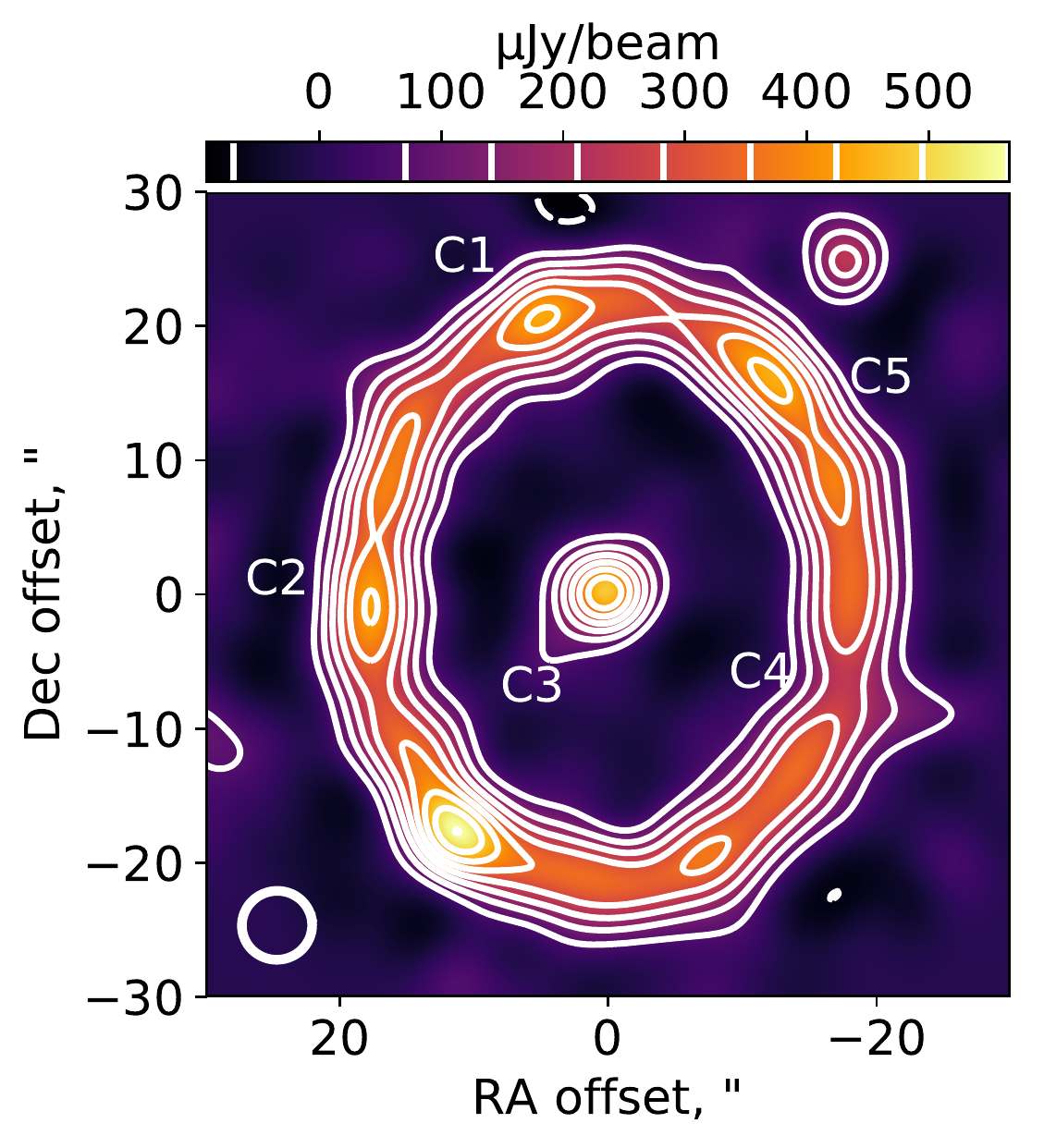}
	\caption{Restored image of the MCMC best-fit model smoothed with a 5\arcsec{} Gaussian in order to highlight the structure in the disc. Contours are in multiples of $\pm2\sigma$. Deviations from the MCMC model identified from the residual plots of \autoref{fgres} that are discussed in the text are labelled. 
	}
	\label{fcleanmcmc}
\end{figure}

In the lower resolution residual images shown in Figure \ref{fgres}, we clearly see a number of significant residuals coincident with the belt. In the map smoothed with a 2.5\arcsec{} Gaussian, we see three $>3\sigma$ residuals in the north, east and northwest parts of the belt, as well as one $<-3\sigma$ residual in the southwest part of the belt and even some low level emission between the belt and the star in the southeast. The features to the east and northwest also show up as $>3\sigma$ residuals in the map smoothed with a 5\arcsec{} Gaussian. Given the extended nature of these features, it is unclear whether the lower significance of the northern overdensity and southwestern underdensity in the 5\arcsec{} image compared to the 2.5\arcsec{} refutes their existence or demonstrates that their scale is smaller than that of the eastern and northwestern residuals. All these features are labelled in Figure \ref{fcleanmcmc}, which shows the restored image\footnote{I.e. the best-fit model convolved with the \clean{} beam and added to the residuals.} after smoothing with a 5\arcsec{} Gaussian.

\section{Resonant clumps}
\label{sclumps}
In Section \ref{sdisc} we identified extended residuals around the main belt implying that dust is more concentrated in certain locations of the disc than others. As discussed in the introduction, similar clumps have been identified in previous observations of \epseri{}, although some have been identified as background galaxies and the low significance of others means that their existence remains contested. Assuming they are real, the leading theory to explain these clumps is that an outward-migrating planet has swept exterior debris into resonances \citep{wyatt06, krivov07}, which also simultaneously explains the disc edge locations.
In this scenario, as the hypothetical planet migrates, the nominal locations of its mean-motion resonances migrate with it, and if these move through regions of debris then planetesimals can get efficiently trapped into resonance \citep{wyatt03}. If the planet migrates sufficiently far, then the result would be a debris disc with {a significant population of resonant planetesimals, where the semimajor axes of the innermost and outermost planetesimals correspond to the nominal locations of mean-motion resonances (although since such planetesimals would be eccentric, the disc edges may not lie exactly at the nominal resonance locations)}. Such a disc would have azimuthal 
variations arising from resonant structures. Since \cite{booth17} showed that a planet at ${48\au}$ would have 3:2 and 2:1 nominal resonance locations coincident with the disc's FWHM locations (which would be expected for a disc with grains on low eccentricity orbits)
and the disc shows signs of azimuthal features, this migrating-planet scenario warrants further consideration. 

\subsection{$N$-body simulations}
\label{snbody}

{To determine whether a migrating planet could produce disc morphologies similar to that observed for \mbox{$\epsilon$ Eri}}, we ran a suite of \mbox{$n$-body} simulations modelling an outward-migrating planet interacting with an exterior debris disc. Each simulation comprised the star, one planet, and an initially unexcited debris disc. The migration would likely be driven by debris scattering, but since this depends on various unknowns (disc mass, surface density and system { architecture; \citealt{ida00, tsiganis05, kirsh09, bromley11, friebe22})}, we chose to impose an artificial migration rate on the planet, and to use test particles to represent planetesimals. We { forced} the planet's semimajor axis to increase at a constant rate, whilst its eccentricity { remained} zero\footnote{The planet eccentricity is expected to be low for several reasons. First, an eccentric planet would impose eccentric structure on the disc {\citep{pearce14, pearce15}}, and such structure is not observed. Second, dynamical friction during scattering would damp planet eccentricity. Third, non-zero planet eccentricity would decrease the trapping probability in low-order resonances \citep{reche08}, but these are the resonances that we expect to populate in our scenario.}. { This approach gave us increased flexibility to explore the parameter space, because each simulation could be post-processed to probe different disc setups as described below.

{We ran 15 individual simulations, testing planets with masses of 0.02 to ${1\mJup}$, initial semimajor axes of 25 to ${47\au}$, and migration rates of $0.01$ to ${1\auPerMyr}$. We can rule out a planet more massive than ${1\mJup}$ as such a planet should have been detected by \emph{Spitzer} observations \citep{janson15}. A lower limit on the planet mass is less easy to define. \citet{pearce22} showed that a non-migrating planet would require \mbox{mass ${\geq 0.19\pm0.04\mJup}$} and semimajor \mbox{axis ${\leq 52.7^{+0.9}_{-1.0}\au}$} to have ejected non-resonant debris from the region just interior to the disc, however, the planet could be less massive if debris is instead cleared by resonant sweeping and so we test masses as low as 10\% of this value. The choice of migration rates is comparable to the rates that the giant planets are thought to have migrated due to planetesimal scattering in the early Solar System \citep{tsiganis05}.} The debris discs initially spanned 48--80~au, comprising ${10^3}$ to ${10^4}$ particles distributed such that the surface density varied with stellocentric distance $r$ as ${\propto r^{-1.5}}$ (like the Minimum Mass Solar Nebula; \citealt{weidenschilling77, hayashi81}).} Each disc particle had an initial inclination and eccentricity uniformly drawn between 0--$5^{\circ}$ and 0-0.05 respectively, and an argument of pericentre, longitude of ascending node and mean anomaly each uniformly drawn between 0--${360^{\circ}}$ { as in \cite{pearce21}}. The simulations were performed with {\sc rebound}, using the {\sc ias15} integrator \citep{rein12, rein15}{, and we ran each simulation until the planet reached at least ${48\au}$ (i.e. until the nominal location of the 2:1 resonance reached the outer edge of the observed disc).

Although we only ran 15 $n$-body simulations varying three parameters, since our simulations used test particles and fixed planet-migration rates, the particles themselves did not affect the overall system dynamics. This allowed us to consider different initial disc setups by varying the disc parameters in a post-processing step, without having to run a new simulation for each setup. 
We considered discs with inner edges ranging from 48 to ${76\au}$ in ${1\au}$ steps, with widths of 1 to ${28\au}$ in ${1\au}$ steps (noting that the outer edge could not exceed ${80\au}$ due to the simulation setup). For each disc configuration we constructed an image by rotating the simulated system to a similar orientation to the observed disc on the sky (sometimes using a slightly different orientation, discussed below), scaling the particles for emission (${\propto r^{-1/2}}$), and convolving the result with a 2D Gaussian
to produce an image comparable to Figure \ref{fcleanmcmc}. We varied the position angle of the planet when constructing the images, and also tested an alternative initial disc surface density profile of ${\propto r^{-1}}$ (versus ${\propto r^{-1.5}}$ in the original simulations) by re-weighting the brightnesses of individual particles.

We focus on reproducing the azimuthal surface brightness profile, because this is a key prediction of the resonant trapping scenario which is motivated by the observation of clumps, but we also consider the disc location, width and shape in determining {our preferred model. We assume that the dust seen by ALMA has the same morphology as the underlying planetesimals (this assumption will be addressed in the following section). In order to ease the comparison between the models and the observation, we use a least-squares fitting technique to find the parameters that produce an azimuthal profile that is most similar to the observed profile.} For each image, we calculate the azimuthal surface brightness profile by dividing the image into 100 radial sectors, and finding the peak surface density in each sector (we mask the region between position angle 132 and ${162^\circ}$, which corresponds to the southeast point source). We construct an equivalent profile from the image in Figure \ref{fcleanmcmc}, where we also omit flux from within ${32\au}$ and outside ${86\au}$ in projection (to omit the star and northwest point source). Both profiles are normalised to 1, and compared to each other using a {least-squares}
fit. We repeat this process for all system setups, to identify the setups that well-reproduce the observed azimuthal brightness profile. {However, we also} manually examine these simulations, removing any where the disc location, width or shape is very dissimilar to the observation. 

Figure \ref{fig: simAllPlots} shows our {preferred model that results from this process. In this simulation}, a ${0.2\mJup}$ planet has migrated outwards from ${37\au}$ at ${0.2\auPerMyr}$ for ${24\myr}$, while interacting with a disc with initial span 61 to ${64\au}$ and initial surface density ${\propto r^{-1.5}}$. In this particular configuration the disc is entirely comprised of particles in the 2:1 resonance. The resonance was initially located interior to the disc inner edge, and swept up planetesimals as the planet moved outwards. During this interaction, the semimajor axes of resonant debris remained close to the (migrating) nominal resonance location, whilst their eccentricities librated up to 
some ever-increasing maximum\footnote{If planet migration stalls, then the maximum eccentricities of resonant particles stop increasing.}. 
The particles in the 2:1 resonance form a structure with two broad clumps centred approximately ${100^\circ}$ ahead of and behind the planet, which arise through the splitting of the 2:1 resonance into upper and lower variants (discussed below; see also \citealt{wyatt03}). Two narrow clumps also form at 0 and ${180^\circ}$ from the planet, caused by the overlap of the upper and lower 2:1 resonance structures. The result is a debris morphology that broadly reproduces the observed disc. If this model is correct, then the observed clump C2 is really the peak of a broad clump comprising material in the 2:1 resonance, and likewise C5 is part of the other broad 2:1 clump. The observed clump C1 arises through the overlap of upper and lower 2:1 structures.
The underdensity C4 marks the azimuthal edge of one 2:1 clump, and a corresponding underdensity may also exist in the southeast part of the disc (currently co-located with the background point source). We find that not only the morphology of the clumps, but also the ratio of the clump peak flux to the inter-clump flux matches that seen in the observations.

Notably, this simulation does not include particles in the 3:2 resonance. As previously noted, \citet{booth17} predicted (by analogy with the classical Kuiper belt) the existence of a planet at 48~au, because such a planet would produce 3:2 and 2:1 resonances with nominal locations coincident with the inner and outer edges of the observed belt. However, we found that our simulations that trapped material in both the 3:2 and 2:1 resonances ended up producing belts much broader than that observed. This is because the particles in these resonances must have eccentric orbits to produce visible clumps, so particles with semimajor axes equal to a nominal resonance location will have orbits that extend inwards and outwards from this point. We also found that simulations
with significant 3:2 debris populations generally produced worse fits to the observations, because
clumps arising from the 3:2 resonance are centred at ${\pm 90^\circ}$ from the planet, which does
not match the azimuthal locations of the observed clumps. The lack of material in the 3:2 resonance occurs naturally in our simulations where the 2:1 resonance is initially interior to the inner edge of the disc. As the planet migrates, the strong 2:1 resonance would typically capture almost everything it encountered and leave very little for the 3:2 resonance.

\begin{figure*}
  \centering
   \includegraphics[width=17cm]{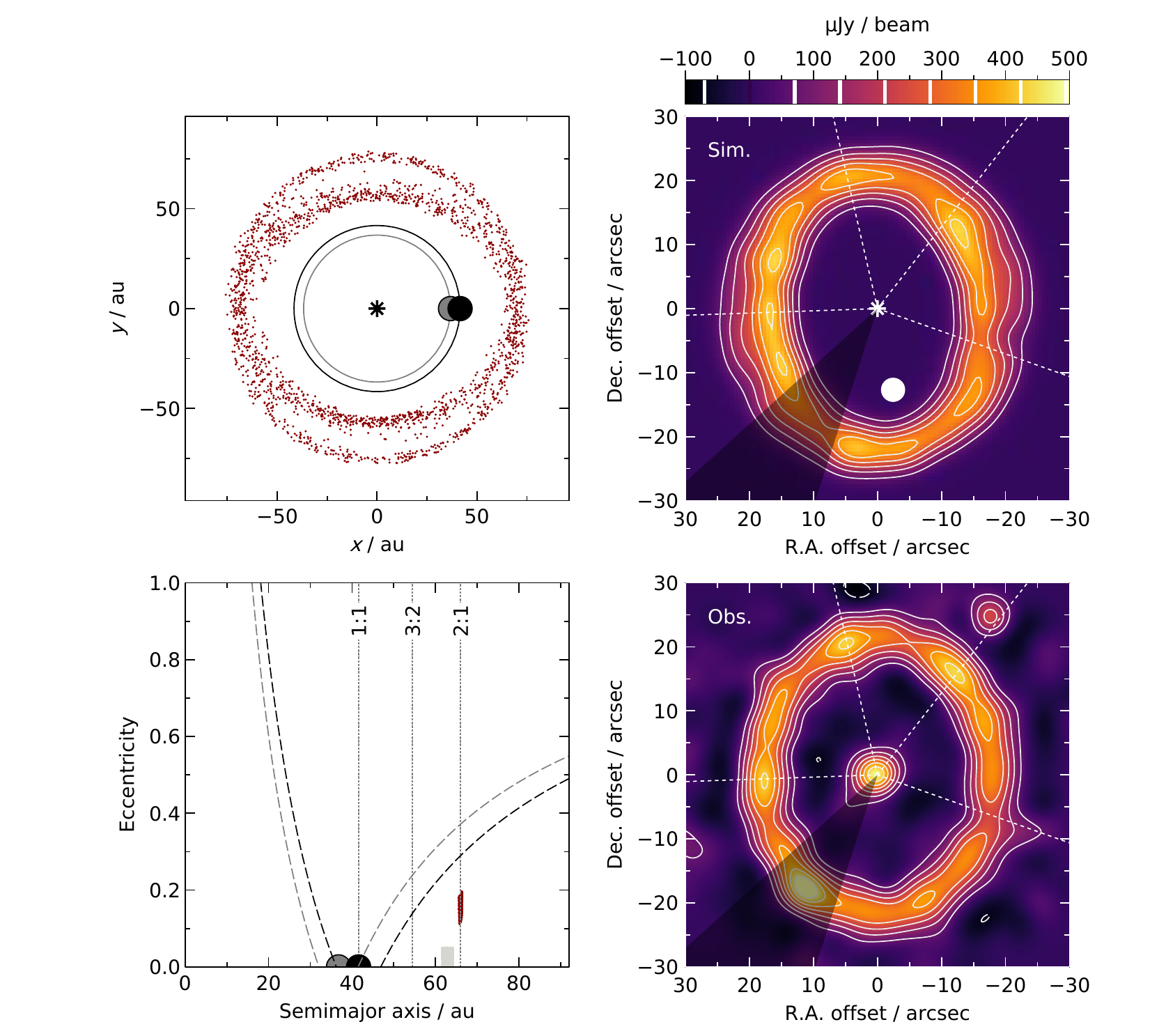}
   \caption{ Simulation of a hypothetical planet migrating outwards into the \mbox{$\epsilon$ Eri} disc (Section \ref{snbody}). The simulation has a ${0.2\mJup}$ planet migrating from 37 to ${42\au}$ at ${0.2\auPerMyr}$ for ${{24}\myr}$, into a disc with initial span 61 to ${64\au}$ and initial surface density ${\propto r^{-1.5}}$. The disc comprises 2300 particles. Top-left panel: face-on view of the simulation, where the \mbox{${x,y}$ plane} is the disc midplane. The points show planetesimal positions and the asterisk is the star. The large circles and lines show planet positions and orbits respectively, at the simulation start (grey) and end (black). Top-right panel: simulated image, where the system has been rotated to a similar orientation to the \mbox{$\epsilon$ Eri} disc on the sky (with a ${10^\circ}$ smaller inclination), scaled for emission, and convolved with a 2D Gaussian with FWHM ${5.2\arcsec}$. The resulting image was then further scaled such that the peak brightness of the eastern clump (C2) matches the observation. The hypothetical planet (white circle) is at position angle $190^\circ$ and orbits in an anti-clockwise direction. Bottom-left panel: particle eccentricities and semimajor axes. The shaded grey region shows the particles at the initial time. Dashed lines show particle orbits that could come within \mbox{3 Hill} radii of the planet at the initial (grey) and final (black) times. Dotted lines show the nominal locations of mean-motion resonances at the final time. Bottom-right panel: observed disc from
Figure \ref{fcleanmcmc}. White dotted lines on the right-hand panels mark the azimuthal
locations of observed features for direct comparison between the simulation and observation, and the shaded region between position
angles 132 and 162$^\circ$ corresponds to the location of the southeast point source in the observations, which is probably a background object.}
   \label{fig: simAllPlots}
\end{figure*}

Whilst the simulation in Figure \ref{fig: simAllPlots} reproduces the observed
disc reasonably well, the fit could still be better. The clumps
are not quite as bright as those observed, and the azimuthal
profile differs somewhat on the east side of the disc. We also
note that the simulated disc has a faint secondary structure
extending beyond the main belt and widest at $\pm90^\circ$
from the planet, which is not seen in the observations; whilst the
addition of simulated background noise would reduce the significance of this feature, it is unlikely to remove it entirely.

To produce the simulated image in the top-right panel of Figure \ref{fig: simAllPlots}, we assumed a disc inclination that is ${10^\circ}$ smaller than that inferred directly from the observed disc (${23.7^\circ}$ rather than ${33.7^\circ}$). The reason is that our simulated resonant disc is not exactly circular, but instead slightly elliptical. This can be seen on the top-left panel of Figure \ref{fig: simAllPlots}, where the disc inner edge is closer to the star at ${\pm 90^\circ}$ from the planet than at 0 and ${180^\circ}$. However, since the star is still roughly central with respect to the resonant disc, on the sky the squashed disc would appear as a roughly circular disc that is slightly more inclined than it actually is. We typically find that reducing the simulation inclinations by ${10^\circ}$ results in axial ratios that are more similar to the observations, so if the ${\epsilon \rm \;Eri}$ disc really is resonant, then it may be slightly less inclined than currently thought. Depending on the specific resonances present, we may also expect a small stellocentric offset, although this is ${\lesssim 1\percent}$ for the simulation in Figure \ref{fig: simAllPlots}. Finally, another
consequence of the resonant disc being elliptical is that the
apparent position angle of the observed disc would depend
on planet position, because the elliptical structure is aligned
with the planet. This means that, if the clumps in the \epseri{} disc are caused by material in resonance with a planet,
then we might expect the disc's apparent position angle to
change over time (in addition to the clump locations).

Our best-fitting simulations have the planet at position angles of ${\sim 190^\circ}$ (within a few tens of degrees either side), but a degeneracy in the resonant dynamics means that planet position angles around ${\sim 10^\circ}$ can also reproduce the observed disc. For the 2:1 resonance, if trapped particles have sufficiently high eccentricities then the resonance splits into upper and lower variants, which have libration centres (and hence clumps) either side of ${180^\circ}$ from the planet (see Figure 8 in \citealt{wyatt03}). If particle eccentricities are increased, then these clumps symmetrically move away from ${180^\circ}$. This leads to a degeneracy in reproducing clumps C2 and C5 (on the north side of the disc). If 2:1 debris has smaller average eccentricities, then the 2:1 clumps would lie either side of ${180^\circ}$ from the planet; in this case, to match the observed clump locations the planet would have position angle ${\sim 190^\circ}$. Conversely, if debris eccentricities were larger, then the 2:1 clumps would be closer to 0 than ${180^\circ}$ from the planet; in this case, the planet would have position angle ${\sim 10^\circ}$. Since a migrating planet increases the eccentricity of resonant debris as it migrates, if the planet migrates a shorter distance then the C2 and C5 clumps would be more consistent with a planet at position angle ${\sim 190^\circ}$, but if the planet migrated further then these clumps would imply a planet at ${\sim 10^\circ}$. It is difficult to use clump C1 to break this degeneracy, because we found that similar overdensities both aligned and anti-aligned with the planet can be formed by 2:1 resonant particles (and also by higher-order resonances). Also, whilst our best-fitting simulations do not include material in the 3:2 resonance (non-zero particle eccentricities mean the simulated discs become too wide if both the 3:2 and 2:1 resonances contain significant debris), a similar degeneracy would arise if the 3:2 resonance were populated; clumps form from 3:2 resonant particles at ${\pm90^\circ}$ from the planet, so a disc with a large 3:2 resonant population would also look similar if the planet position angle were changed by ${180^\circ}$. So if the ${\epsilon \; \rm Eri}$ disc is resonant, then we predict the responsible planet to lie either at position angle ${\sim 190^\circ}$ or ${\sim 10^\circ}$. A planet at these locations would be well-separated from the star in projection, which is favourable for future planet searches.

The simulation on Figure \ref{fig: simAllPlots} reproduces the observations reasonably well, but we stress that this is not a unique solution and that many degeneracies arise from the large number of unknown parameters (initial disc location and width, initial planet location, mass and migration rate). We found it possible to fit the disc with alternative resonant scenarios, for example different resonances, initial disc locations, and migration timescales. We also note a degeneracy between planet mass and migration rate; debris can be trapped into resonance just as efficiently by a high-mass, rapidly migrating planet as by a low-mass, slowly migrating planet. In fact, we did not see a strong dependence of disc morphology on planet mass or migration rate; instead, the factors with the largest influence appear to be the \textit{distance} over which the planet migrates, and the initial edges of the pre-interaction disc. We also emphasise that we have not performed a detailed parameter space exploration; whilst post-processing simulations allowed us to explore a broad range of initial disc setups, we only probed 15 combinations of planet initial location, mass and migration rate. It is therefore very possible that the disc could be better fit by some untested system setup, and a more detailed parameter space exploration is required to further explore this interaction.

We showed that a planet migrating outwards could potentially reproduce the observed debris disc. However, we have not attempted to constrain the migration mechanism. If migration were driven by planetesimal scattering, then outward migration would be difficult for a single planet, because the tendency is for the planet to eventually eject unstable debris and migrate inwards  {\citep{kirsh09, friebe22}}. Outward migration is easier {in} a multi-planet system, because an outer planet can scatter material inwards to another planet, which then ejects it. This would cause the outer planet to migrate outwards into the disc, whilst the inner planet migrates inwards\footnote{This process is thought to have driven early inward migration of Jupiter and outward migration of Saturn, Uranus and Neptune in the Solar System \citep{tsiganis05}.}. A possible result would be that much of the debris scattered inwards by the outer planet would since have been ejected by the inner planet(s). This would be compatible with our preferred simulation. Whilst the simulated configuration on Figure \ref{fig: simAllPlots} has no material initially close to the planet, in reality such material should have existed to drive the outward migration, but could have been rapidly ejected by any inner planet(s). We know that a Jovian-mass inner planet resides at several au in the \mbox{$\epsilon$ Eri} system \citep{llop21}, but since our predicted disc-sculpting planet would lie at ${\sim 50\au}$ and have low eccentricity, it is unlikely that this predicted planet could scatter significant quantities of material directly to the inner planet for ejection. Therefore, if the \mbox{$\epsilon$ Eri} disc has been sculpted by an outward-migrating planet, then this would likely be facilitated by one or more additional planets residing interior to our predicted planet; either these additional planets eject scattered debris themselves, or they form a chain that can pass debris right into the inner system (where the known planet can then eject it, e.g. \citealt{bonsor12a}). This may also contribute to the dust in the inner region.

\subsection{Dust in the resonant clumps}
In the above model it is the planetesimals that are trapped in resonance and so one important question to ask is whether the dust remains in the same location as the planetesimals since it is small grains of around the same size as the wavelength of the observations that our observations are most sensitive to. \citet{wyatt06} investigated how dust is lost from resonances due to the increase in eccentricity from radiation pressure.
They derived a formula for the critical diameter below which dust is lost from the resonance, which (assuming large\footnote{In this case, large means grains for which the radiation pressure efficiency is approximately equal to 1, which is true for grains of diameter $\gtrsim20\um$ in the \epseri{} system \citep{wyatt99}.} grains) can be written as
\begin{equation}
 D_{\rm{crit}}\approx3.25\!\times\!10^4\!\left(\!\frac{\rho}{\kgPerMCubed}\!\right)^{\!-1}\!\frac{L_\star}{\lSun}\!\left(\!\frac{M_\star}{\mSun}\frac{M_{\rm{pl}}}{\mJup}\!\right)^{\!-0.5}\!\um,
 \label{edcrit}
\end{equation}
where $\rho$ is the density of the grains, $L_\star$ and $M_\star$ are the luminosity and mass of the star and $M_{\rm{pl}}$ is the mass of the planet. For \epseri{}, \mbox{$L_\star=0.3808\pm0.0015\lSun$} \citep{gaia18} and $M_\star=0.82\pm0.05$~M$_\odot$ \citep{baines12}. We do not have any measure of the density of the grains in the \epseri{} system, but can assume that they are similar to the grains released from comets within our own Solar System. For example, observations of the dust released from comet 67P/Churyumov-Gerasimenko have derived a density of $\sim800\kgPerMCubed$ \citep{fulle17,kwon22}. We can use this value as a rough estimate, although note that depending on the exact composition and porosity, the density of the grains in the \epseri{} system could be up to a factor of $\sim$5 higher or lower than this \citep{guttler19}. For a planet of mass $0.2 \mJup$, as in the example simulation shown in Section \ref{snbody}, we find that $D_{\rm{crit}}\approx40$~$\mu$m. Since the emission detected at a given wavelength is expected to be dominated by grains of a size approximately the same as that wavelength \citep[e.g.][]{backman93}, the millimetre dust that we are seeing is much larger than this critical diameter and so is too large to be removed from resonance by radiation pressure.

Although radiation pressure is not expected to impact millimetre grains, the effect of velocity dispersion could be much stronger.
This effect arises from the slight differences in velocity that the products of a collision
have relative to the centre of mass of the two colliders. These velocities increase the libration
amplitude of the resonant argument and, at sufficiently high velocities, the debris particles
will leave the resonance. As a criterion of staying in the resonance,
one can require that the libration amplitude does not grow
above a certain threshold value $A$. In order to investigate this for the \epseri{} system, \citet{krivov07} conducted numerical integrations of the 3:2 resonance. Although we find that the 2:1 resonance is dominant in our preferred $n$-body simulation (see Section \ref{snbody}), we expect the results to be similar. 
Based on their simulations, 
\citet{krivov07} determined the maximum relative velocity $v_\mathrm{crit}$
that does not break resonance trapping, to be
\begin{equation}
  \frac{v_\mathrm{crit}}{v_\mathrm{k}}
  =
  {\cal A}
  \left(
    \frac{M_\mathrm{pl}/\mJup}
         {M_*/\mathrm{M}_\odot}
  \right)^{\cal B},
\end{equation}
where $v_\mathrm{k}$ is the circular Keplerian velocity
(3.4~km~s$^{-1}$ for the 3:2 resonance at $63$~au),
and the fitting coefficients are
${\cal A} = 0.007 (\pm40\%)$, ${\cal B} = 0.28 (\pm10\%)$ for $A=30^\circ$.
Using $M_\mathrm{pl}=0.2 \mJup$
as in the example simulation, we find
$v_\mathrm{crit} = 16$~m~s$^{-1}$. We note that an order of magnitude change in the assumed mass of the planet results in a factor of two change in the critical velocity.

In order to estimate the typical ejection velocity of collision fragments, we can make use of the analysis in \citet{wyatt02}. We start by assuming that the kinetic energy is distributed evenly amongst all fragments except
for the largest remnant and that the collision fragments are largely produced via barely catastrophic collisions, because lower energy collisions are cratering collisions that produce few fragments and higher energy collisions are rarer since they require larger impactors. This means that the specific energy of the collision is equivalent to the dispersal threshold ($Q_{\rm{D}}^*$) and the mass of the largest remnant is half the mass of the target. We can then combine their equations 6 and 22 to get
 \begin{equation}
  v_\mathrm{ej}=\sqrt{\frac{2f_\mathrm{KE}Q_\mathrm{D}^*}{1+2Q_\mathrm{D}^*/v_\mathrm{col}^2}},
  \label{evej}
 \end{equation}
where $f_\mathrm{KE}$ is the fraction of the impact energy that is imparted to the resulting collisional fragments and $v_\mathrm{col}$ is the impact velocity of the collision. Typically, $2Q_\mathrm{D}^*/v_\mathrm{col}^2\ll1$ and so Equation \ref{evej} can be approximated as
\begin{equation}
 v_\mathrm{ej}\approx\sqrt{2f_\mathrm{KE}Q_\mathrm{D}^*}.
\end{equation}
The value of $f_\mathrm{KE}$ can be inferred from collisional experiments and amounts to a few per cent to several tens of per cent for small debris
\citep[e.g.][]{asada85,hartmann85}, therefore we use $f_\mathrm{KE} \sim 0.01$--0.1 to give an example of typical values, although we note that there is a lot of uncertainty on this. For grains $\sim1$~mm in size, laboratory measurements show a wide range of dispersal thresholds from 10$^2$--10$^4$~J~kg$^{-1}$ \citep[see][and references therein]{leinhardt09}. From this we estimate that the typical ejection velocities are between 1 to 30~m~s$^{-1}$.

Ejection velocities have also been measured in impact experiments directly, with the aid
of high-speed cameras. Values between $\sim 1$~m~s$^{-1}$ and $\sim 35$~m~s$^{-1}$, consistent with the above estimate,
have been reported over a broad range of impactor and target masses, materials,
and collisional velocities
\citep[e.g.,][]{fujiwara86,nakamura91,giblin98,cintala99}.

Comparing these experimentally derived values of $v_\mathrm{ej}$ to the threshold $v_\mathrm{crit}$
inferred from simulations, we conclude that a significant fraction of collisional
debris, at least for the millimetre-sized grains seen by ALMA and assuming a $0.2 \mJup$ planet,
is likely to have $v_\mathrm{ej} < v_\mathrm{crit}$ and, therefore, our model does predict clumpy structure at millimetre wavelengths as well as in the planetesimals.\footnote{\citet{pearce21} reached the opposite conclusion for Fomalhaut, where grains
released from resonant clumps would shear into a smooth ring. This
arises because Fomalhaut b has an extreme orbit, resulting in complex
resonant behaviour that is very sensitive to small changes in debris
orbits (their Figure 7). For \epseri, the less extreme planet orbit
makes resonant debris more resistant to minor velocity kicks.} 

The problem of velocity dispersion becomes more important for lower mass planets and so the presence of dust clumps does imply that the planet mass is above some critical value, however, the large uncertainties on many of the parameters mean that we can only strictly rule out planet masses less than $2\times10^{-5}\mJup$ (assuming the most extreme values for the uncertain parameters, i.e. ${\cal A}=0.01$, ${\cal B}=0.3$, $f_\mathrm{KE}=0.01$ and $Q_\mathrm{D}^*=10^2$~J~kg$^{-1}$).

\subsection{Rotation of the clumps}
So far in this section we have demonstrated that the migrating planet scenario can broadly reproduce the observed features of the disc. However, this alone is not enough to conclusively prove that the features we see are resonant clumps. To do that we need to observe rotation of the clumps in multi-epoch datasets. Such a comparison has been made before by \citet{greaves05}. By comparing their 2nd epoch SCUBA data to the 1st epoch data from \citet{greaves98}, they showed that some of the features remained at the same right ascension and declination, including the bright clump to the east \citep[also clearly identified as a background source by][]{chavez16} and a clump to the southwest (that is consistent with the point source now seen to the southeast of the star in our data -- see Section \ref{spoint}). However, other features (albeit, features with an SNR<3 -- specifically peaks to the northwest, northeast and southeast) were found to move with the star and showed tentative signs of an anti-clockwise rotation of $\sim$1\degr\,yr$^{-1}$, consistent with that expected if the clumps are due to material trapped in MMRs with an interior planet \citep{ozernoy00, quillen02}. \citet{poulton06} undertook a more thorough analysis of the SCUBA datasets and measured a much higher rotation rate of $\sim$2.75\degr\,yr$^{-1}$, although noted that much smaller rotation rates are still consistent with the data.

The expected rotation rate of the clumps is dependent on the Keplerian velocity of the planet as the clumps move with the planet. For a planet with semi-major axis $a_{\rm{pl}}$ around a star of mass $M_\star$, the angular velocity is:
\begin{equation}
 \omega = 360 \sqrt{\frac{M_\star}{\mSun}\left(\frac{a_{\rm{pl}}}{\rm{au}}\right)^{-3}}\,\degr\,\rm{yr}^{-1}.
\end{equation}
Therefore a planet with a rotation rate of $\sim$2.75\degr\,yr$^{-1}$ \citep[as in][]{poulton06} would require the planet to have a semi-major axis of 24~au -- far too far from the disc for it to trap planetesimals into major MMRs. Whereas a planet with a semi-major axis of $a_{\rm{pl}}=42$~au (as in the simulation shown in Section \ref{snbody}) results in an expected rotation rate of the resonant clumps of 0.98\degr\,yr$^{-1}$, consistent with the motion reported by \citet{greaves05}. The data presented in this paper was taken 17 years after that presented in \citet{greaves05}. The $\sim3\arcsec$ positional uncertainty on the SCUBA clumps \citep{greaves05} corresponds to $\sim8\degr$ at 70~au, whilst the positional uncertainty for the brightest of the clumps found here (the eastern clump has a significance of 3.8$\sigma$ in the map smoothed with a 5\arcsec{} Gaussian) corresponds to $\sim4\degr$ (based on equation \ref{ealmaposunc}). Therefore a 1\degr\,yr$^{-1}$ rotation rate would not be significantly detected.
In addition, it is important to keep in mind that resonant clumps are not individual points. They are broad concentrations of particles that are individually orbiting at a slower rate than the clump and each clump that we see can potentially be a superposition of particles in different resonances. Given this and the low SNR of the clumps in each dataset, a more in depth analysis is necessary to identify which clumps are the same and determine any possible rotation, which we leave for future work.

\section{Emission in the inner region}
\label{sinner}
\epseri{} is well known to have warm emission close to the star \citep{backman09,greaves14a,su17}. By combining the available mid-infrared resolved images and photometry, \citet{su17} demonstrated that in situ dust production is most likely responsible for the observed emission and this can be reproduced by either a broad disc extending from around 3 to 20~au or two belts at around 2 and 8~au. Using the AzTEC 1.1~mm data, \citet{chavez16} also found that there is evidence for emission between the main belt and the star at long wavelengths, primarily extending to the southeast and northwest of the star, although the low significance of this emission meant that they were not able to constrain its origin.

\begin{figure}
	\centering
	\includegraphics[width=0.48\textwidth]{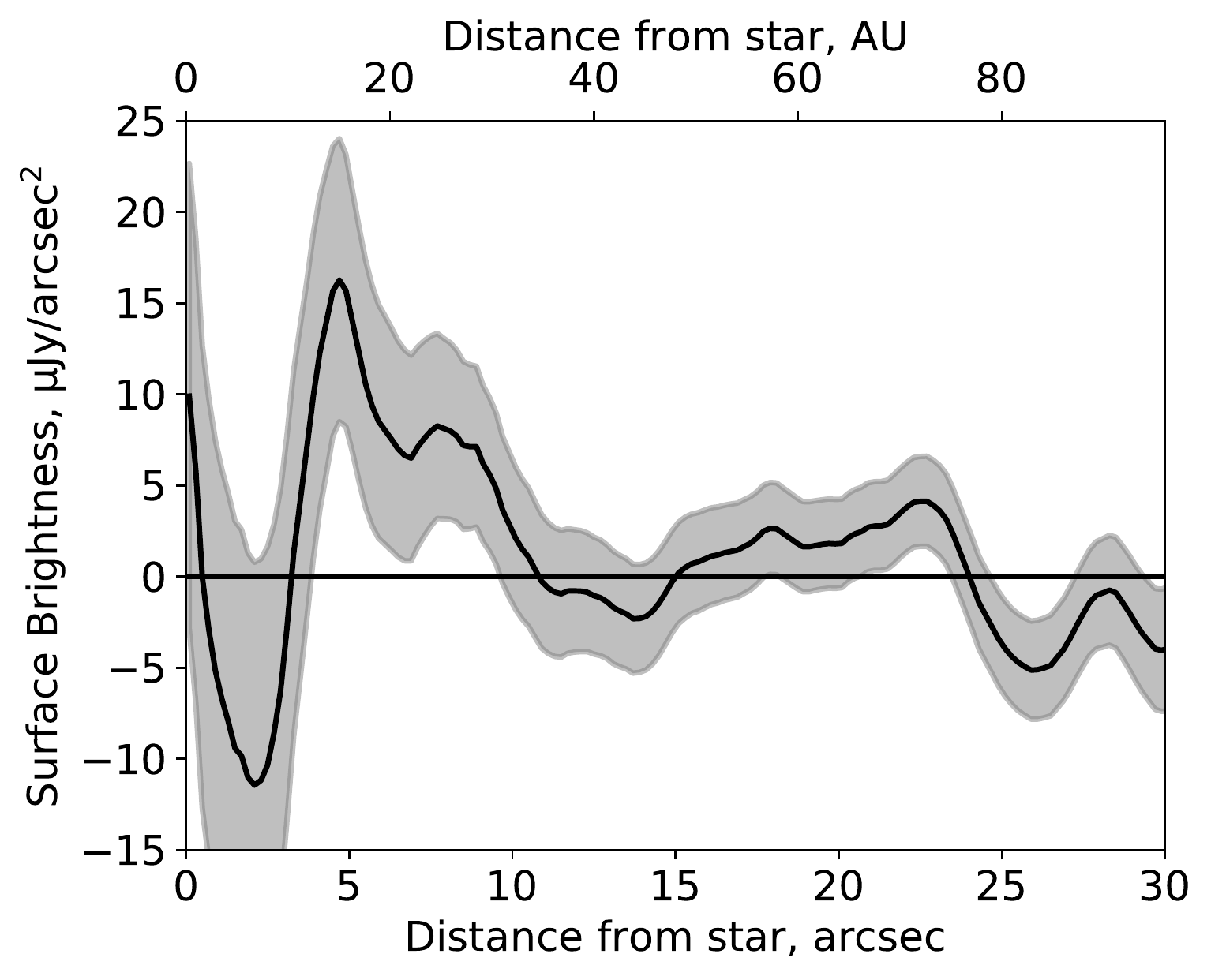}
	\caption{Deprojected and primary beam-corrected surface brightness radial distribution of the residuals shown in the left plot of Figure \ref{fgres} considering only the data between position angles of 120\degr{} and 150\degr{}. The rise in emission between 10 and 20~au reaches a peak of $\sim2\sigma$.}
	\label{fraddist120}
\end{figure}

In our new data we do find a rise in the radial profile between 10 and 20~au (see Figure \ref{fcleanprof}) but the uncertainties are large and so this is not significant. However, there are azimuthal variations in the emission and we see from Figures \ref{fgres} and \ref{fcleanmcmc} that there is some emission to the southeast of the star interior to the main belt (in particular around the position labelled C3), in the same location as that seen in the AzTEC data. Inspired by this, we take the image of the residuals left after subtraction of the best MCMC model (the left plot of Figure \ref{fgres}), divide the image into 30\degr{} sectors, correct the image for the primary beam, deproject and azimuthally average for each sector. C3 is in between position angles of 120 and 150\degr{} and the radial profile for this is shown in Figure \ref{fraddist120}. Here we see that there is a higher rise in emission between 10 and 20~au, although it still only reaches a peak of $\sim2\sigma$.

If the asymmetry of the inner emission that is observed in both our data and the AzTEC data is not just an artifact of the noise, then they may be resonant clumps caught in resonance with a planet much closer to the star than the one responsible for the clumps in the main belt. 
Deeper observations of the inner region are necessary to determine the distribution of dust there and how it may relate to other planets in the system. Observations with other telescopes will also be of benefit. For instance, the \emph{James Webb Space Telescope} (\emph{JWST}) will observe \epseri{} at a number of wavelengths with both MIRI and NIRCam \citep{beichman17}. At 25.5$\um$, these observations will have an angular resolution of 1.0\arcsec{}, close to that of our ALMA observations, and so should be able to pick up similar features. However, note that these mid-infrared observations will be sensitive to small grains around the critical size given by Equation \ref{edcrit} and so any clumps may be smoothed out due to the small grains being pushed out of resonance by radiation pressure \citep{wyatt06}.

\section{Conclusions}
\label{sconc}
In this paper we present ALMA cycle 7 observations of the \epseri{} system. We detect the star, the main belt and two other point sources. We demonstrate that these two point sources are consistent with being background galaxies. We find that a narrow belt consistent with the \citet{booth17} fit to the northern arc works well at fitting the entire belt. After accounting for the background galaxies, we find that significant residuals remain in the region of the belt, as observed in prior observations. The two most prominant residuals (detected at $>3\sigma$ significance) are to the east and northwest of the star. 

Under the assumption that these residuals are resonant clumps, we run a set of $n$-body simulations with which we demonstrate that similar structures can be produced by a migrating planet trapping planetesimals into its 2:1 mean motion resonance. Based on the symmetry of the clumps, we expect this planet to be at a position angle of either ${\sim 10^\circ}$ or ${\sim 190^\circ}$. From the limited parameter space we explore here, we find that the features of the disc are best reproduced by a ${0.2\mJup}$ planet migrating from 37 to ${42\au}$ at ${0.2\auPerMyr}$ for ${{24}\myr}$. However, we note that we are not able to strongly constrain any of these parameters as there are a lot of degeneracies between them and a much larger number of simulations is needed to fully explore parameter space, which we leave for future work. In addition, if a planet is detected -- for instance, \emph{JWST} observations should be able to detect planets down to a mass of $\sim0.3\mJup$ \citep{beichman17} -- then its mass and orbit will help us to break these degeneracies and determine the migration history of the system.

In order to conclusively prove that the observed residuals are due to resonant trapping, a thorough and consistent analysis of all of the observed datasets and continued monitoring is necessary to detect rotation of the clumps.

\section*{Acknowledgements}
The authors thank Wayne Holland for providing the reduced SCUBA-2 data, J\"urgen Blum for discussions about laboratory data, Dirk Petry for discussions about the ALMA primary beam and the referee for their comments on the manuscript. MB, TDP, AVK and TL are supported by the Deutsche Forschungsgemeinschaft (DFG) grants Kr 2164/13-2, Kr 2164/14-2, Kr 2164/15-2 and Lo 1715/2-2. This project has received funding from the European Research Council (ERC) under the European Union's Horizon 2020 research and innovation programme under grant agreement No 716155 (SACCRED). VF acknowledges funding from the National Aeronautics and Space Administration through the Exoplanet Research Program under Grant No. 80NSSC21K0394 (PI: S. Ertel). MC thanks support from CONACyT through grant CB-2015-256961. ALMA is a partnership of ESO (representing its member states), NSF (USA) and NINS (Japan), together with NRC (Canada), NSC and ASIAA (Taiwan), and KASI (Republic of Korea), in cooperation with the Republic of Chile. The Joint ALMA Observatory is operated by ESO, AUI/NRAO and NAOJ. The National Radio Astronomy Observatory is a facility of the National Science Foundation operated under cooperative agreement by Associated Universities, Inc. This research made use of \textsc{astropy}, a community-developed core Python package for Astronomy \citep{astropy13}, and \textsc{corner} \citep{corner}.

\section*{Data availability}
This paper makes use of the following ALMA data: ADS/JAO.ALMA\#2019.1.00696.S and ADS/JAO.ALMA\#2017.1.01644.S. This paper makes use of \emph{Herschel} data with observation ID 1342191177. This paper makes use of data from JCMT, project IDs MJLSD01 and M11BGT01. Data resulting from this work are available upon reasonable request.

\bibliographystyle{mnras}
\bibliography{thesis}{}

\bsp

\end{document}